\documentclass[tighten,twocolumn]{aastex631}

\usepackage{amsmath}
\usepackage{ulem}
\usepackage{xcolor}   
\usepackage{url}

\newcommand{\pja}[1]{#1}

\catcode`_=\active
\newcommand_[1]{\ensuremath{\sb{\mathrm{#1}}}}
\catcode`^=\active
\newcommand^[1]{\ensuremath{\sp{\mathrm{#1}}}}

\let\vec\mathbf

\shorttitle{}
\shortauthors{Judge et al.}

\newcommand{\uat}[2]{\href{http://astrothesaurus.org/uat/#2}{#1 (#2)}}

\newcommand{\hao}{
High Altitude Observatory,
National Center for Atmospheric Research,
Boulder CO 80307-3000,
 USA}

\newcommand{\nso}{
National Solar Observatory, 3665 Discovery Drive,
Boulder CO 80303,
 USA}

\newcommand{\nsomaui}{
National Solar Observatory, 22 \'{}\={O}hi\'{}a K\={u} Street,
Pukalani, HI 96768, USA}

\newcommand{\figvfield}{
\begin{figure*}
\centering
\includegraphics[width=0.8\linewidth]{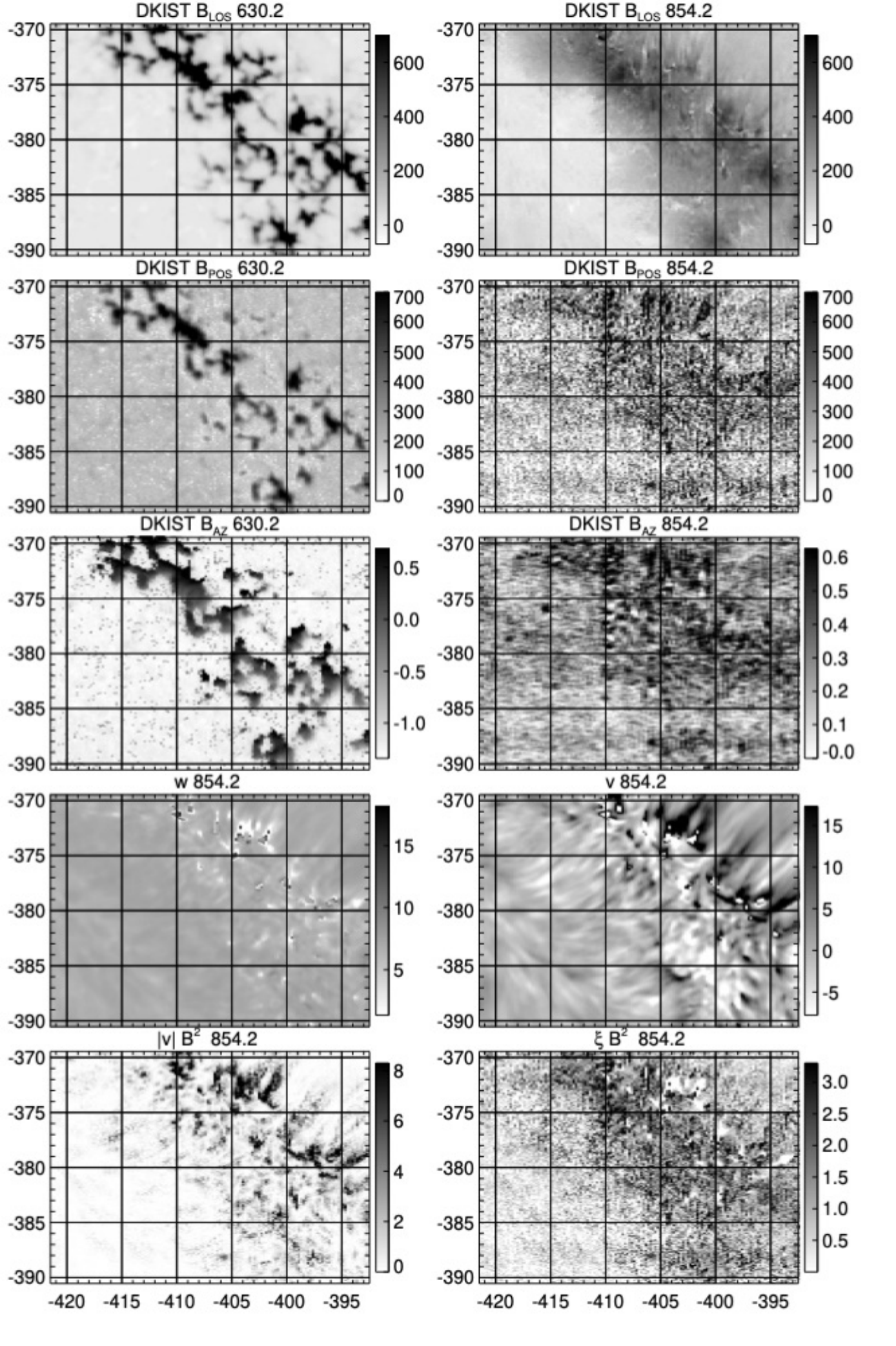} 
\caption{Components of the vector magnetic field derived from the weak field approximation of the 630.2 and 854.2 nm lines are shown, along with RMS velocities and the field amplitude $|B|$.  The bottom panels
show upper limits to the
upward Poynting flux into 
the solar corona.
}
\label{fig:vfield} 
\end{figure*}
}

\newcommand{\fighe}{
\begin{figure*}
\includegraphics[clip, trim=0.cm 2.2cm 0.cm 0cm,width=\linewidth]{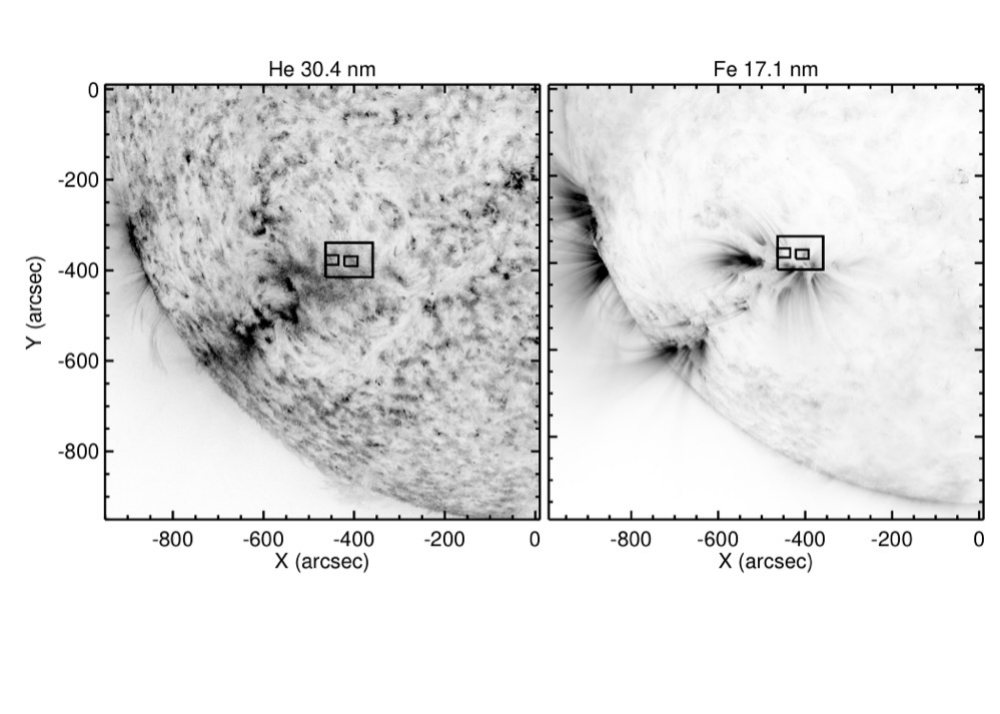} 
\caption{The SE quadrant of the Sun is shown as a sum of images obtained at 30.4 nm (``transition region'') and 17.1 nm (corona) on June 3 2023. An inverse color scale is used. Rectangular regions of interest are shown in more detail in  Figure~\ref{fig:fov}}.
\label{fig:he} 
\end{figure*}
}

\newcommand{\figfov}
{\begin{figure}
\includegraphics[width=\linewidth]{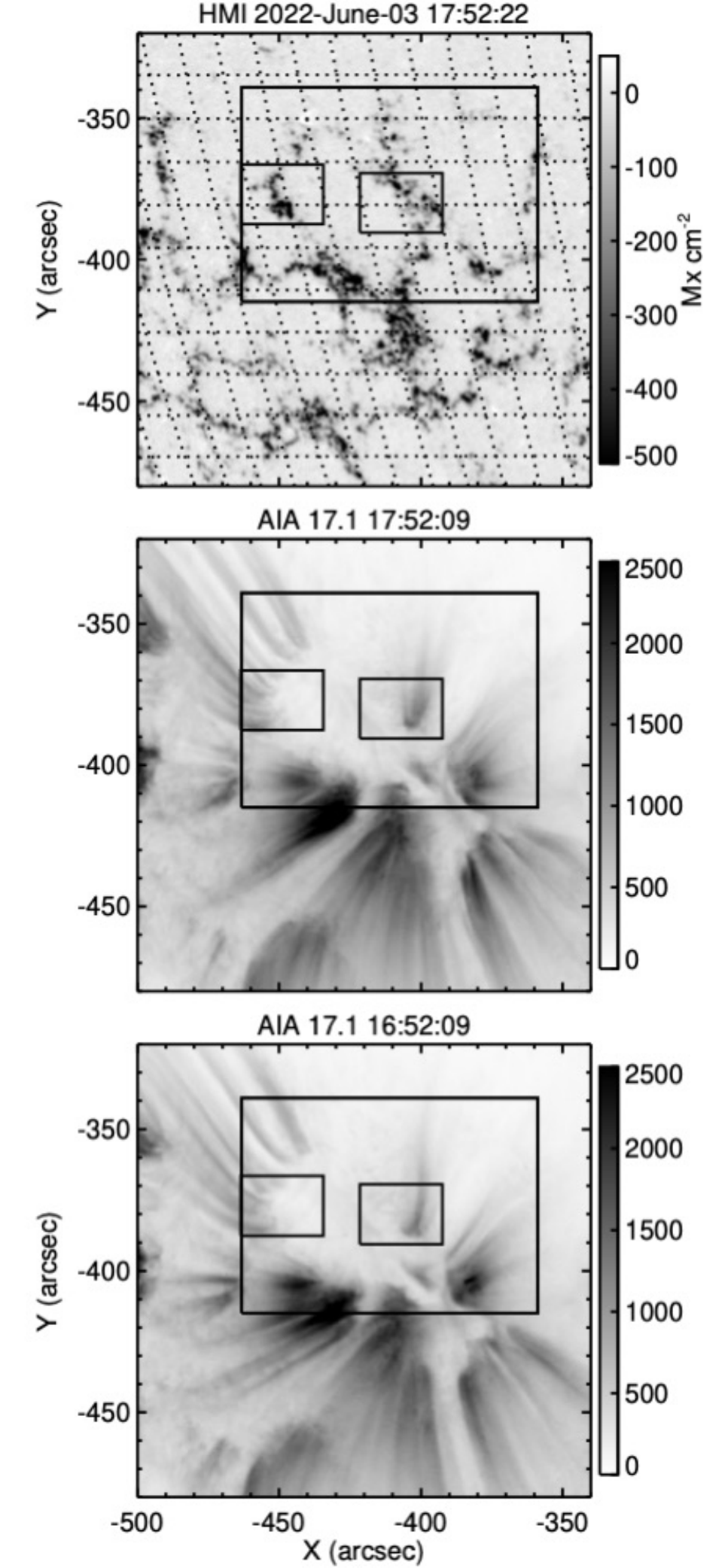}  
\caption{The area observed by arm 1 of
the ViSP is shown by the large black box.
\pja{The images show
line of sight
magnetic fields and coronal data from the SDO
spacecraft. } The 
small boxes show 
regions discussed 
in more detail.  
The AIA data at 17.1 nm 
\pja{were acquired 1 hour apart to illustrate 
the typical evoluton of plasma 
loops, at 16:52 and 
at 17:52 UT on June 3 2022.}
}
\label{fig:fov} 
\end{figure}
}

\newcommand{\figwest}
{\begin{figure*}
\includegraphics[width=\linewidth]{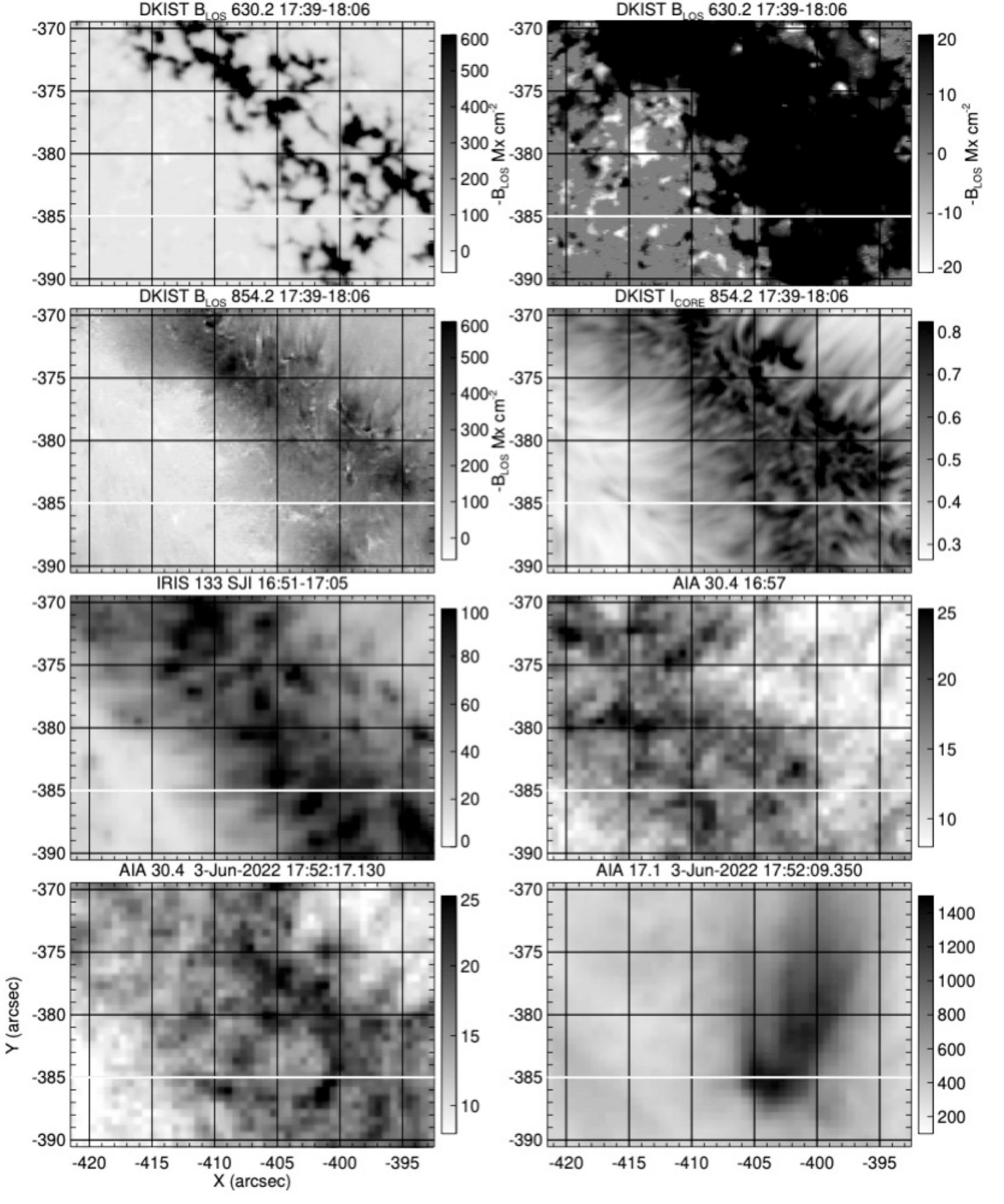} 
\caption{
The western area of interest shown in 
Figure~\ref{fig:fov} is
shown as in Figure~\ref{fig:east}.
Again, data from 16:51 UT (in the third row) have been 
rotated to 17:52 UT, to show evolution of the AIA 30.4 nm channel over 1 hour. Notice the 
correspondence between the coronal footpoint centered at  $X=-403,Y=-385$, and the near- contemporaneous ring-like structures in the core of the 854.2 chromospheric line and 30.4 nm emission line 
at 17:52:17 UT. The white line marks the slice along which the line plots of Figure~\ref{fig:westline} are extracted.   The upper right panel is a saturated 
version of the upper left panel.
}
\label{fig:west} 
\end{figure*}
}

\newcommand{\figeast}
{\begin{figure*}
\includegraphics[width=\linewidth]{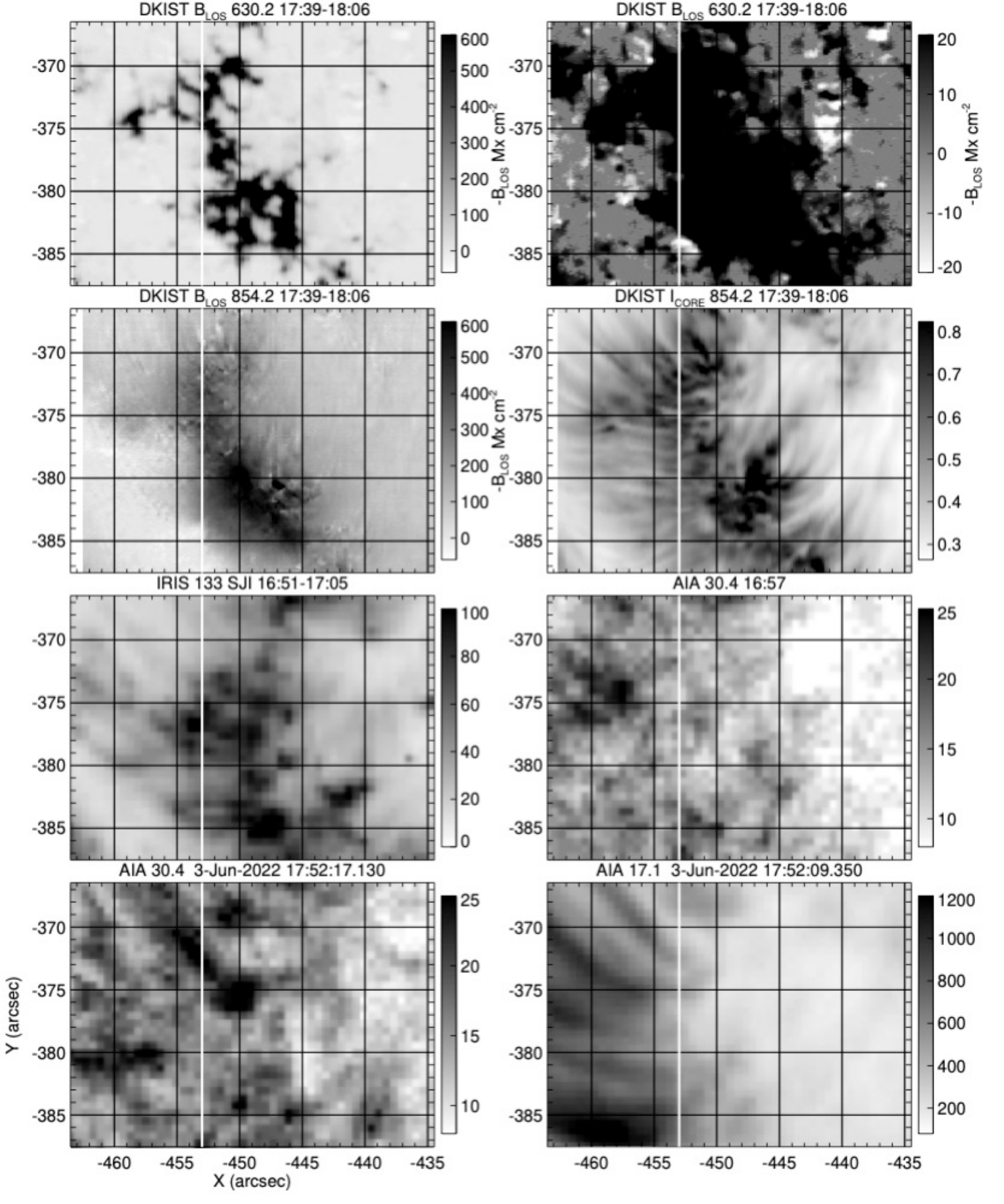} 
\caption{The eastern area of interest shown in 
Figure~\ref{fig:fov} is shown in DKIST data (upper two rows),
slitjaw images at UV 
wavelengths from IRIS (third row), 
and AIA data (lower panels). \pja{IRIS data from 16:51 UT (in the third row) have been 
rotated to 17:52 UT, revealing the 
evolution over 1 hour of the 30.4 nm emission. 
Intensity data are in numbers of counts, except for the ViSP
panel which is 
normalized to a continuum intensity of 
one.}
The white line marks the slice along which the line plots of Figure~\ref{fig:eastline} are extracted.   The upper right panel is a saturated 
version of the upper left panel.
}
\label{fig:east} 
\end{figure*}
}

\newcommand{\figsense}
{\begin{figure}
\includegraphics[width=\linewidth]
{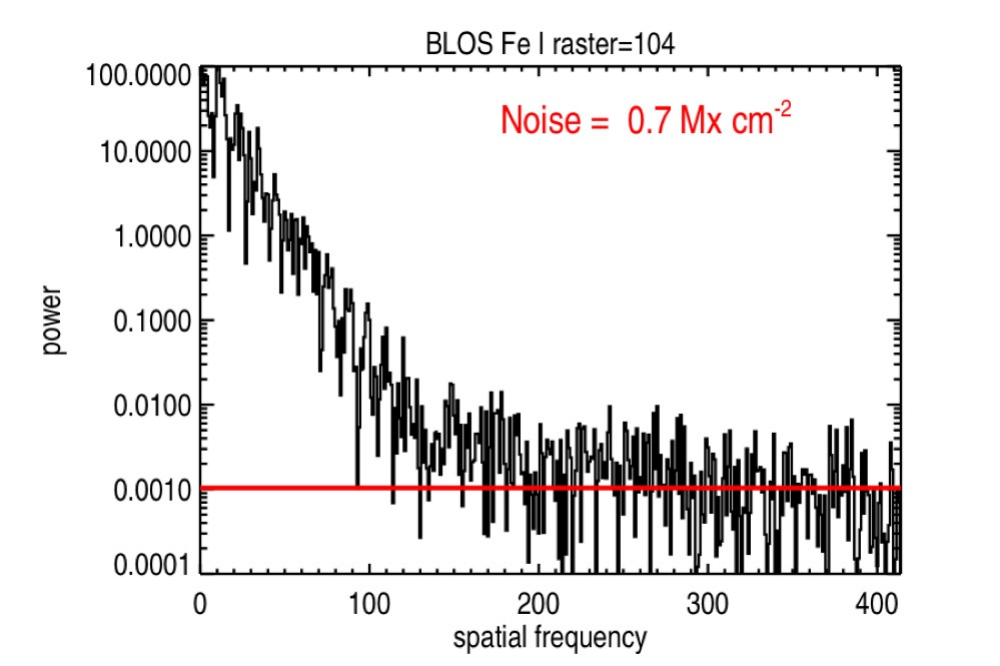} 
\caption{The power spectrum
of spatial variations along the ViSP slit for raster 104.  The
noise is estimated by taking the square root of the sum of power underneath the red line. The point where the red line 
departs from the black near $X=150$
is a measure of the smallest structure
present in the data, whose Nyqvist
sampling frequency is 
${\mathcal N}=0.5/0.059$ arcsec$^{-1}$.  The spatial
resolution along the slit is thus
$\approx 1/(150{\mathcal N}/414) \approx 
0\farcs3$, perhaps $0\farcs2$ for this particular power spectrum, depending
on where the noise is considered as 
flat or ``white''.}
\label{fig:sens} 
\end{figure}
}

\newcommand{\figbrwjv}
{\begin{figure}
\includegraphics[width=\linewidth]{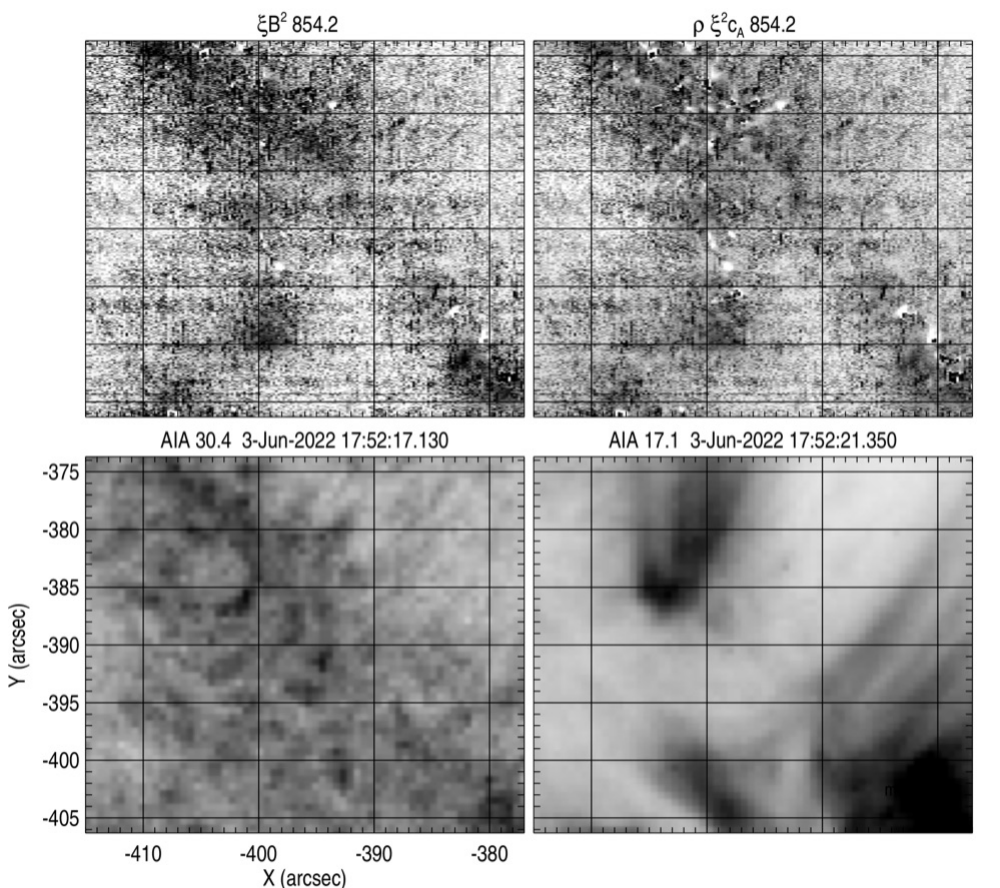} 
\caption{Relative Poynting flux estimates 
and AIA images are shown for dataset AVORO
from 17:52, 3 June 2022.
}
\label{fig:brwjv} 
\end{figure}
}

\newcommand{\figeastline}
{\begin{figure}
\includegraphics[width=\linewidth]{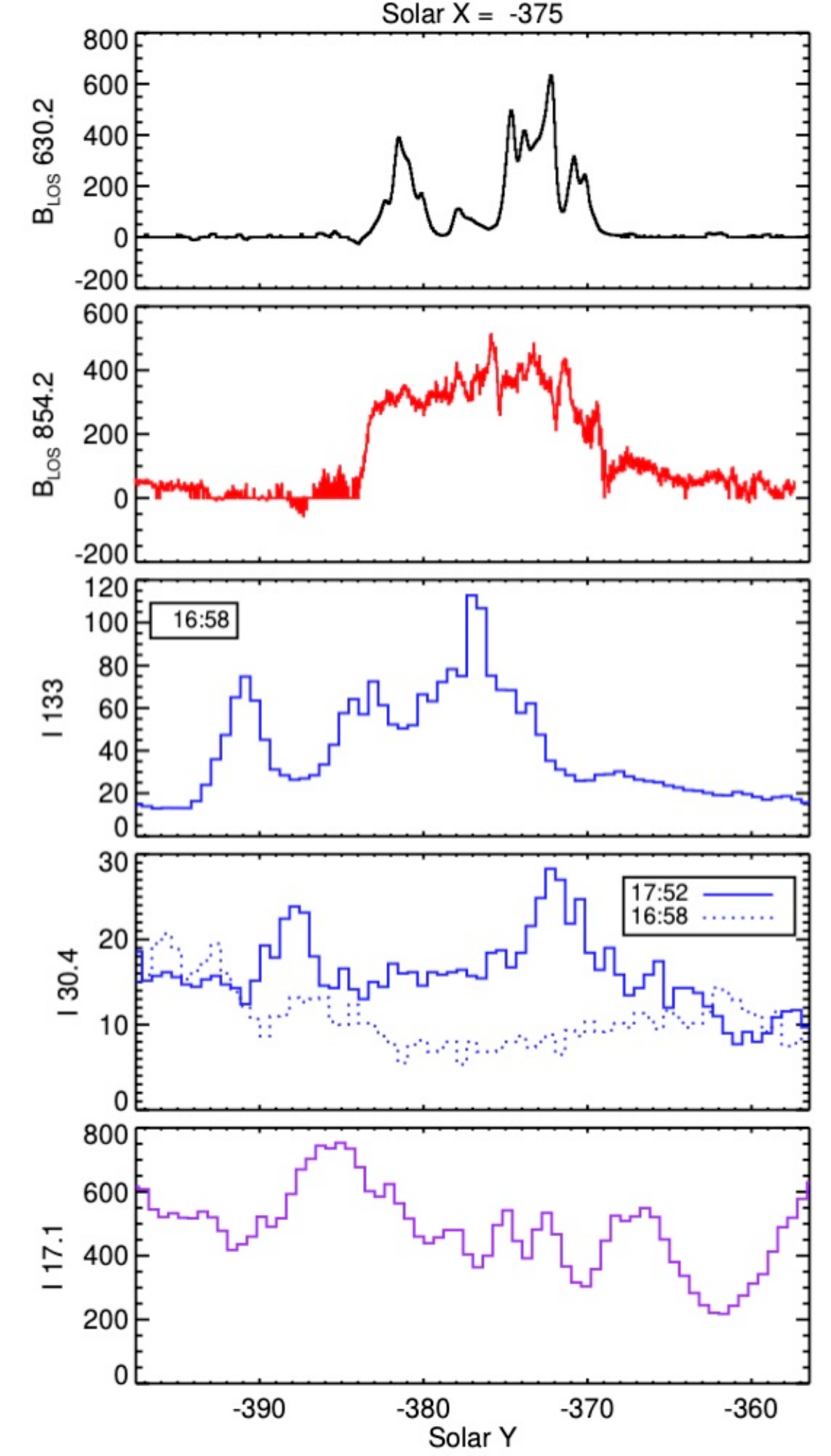} 
\caption{Line plots are shown of  quantities extracted from
the white line of Figure~\ref{fig:east}.
The different colors correspond to
different plasmas (black=photosphere, red=chromosphere, blue = transition region, purple=corona).
}
\label{fig:eastline} 
\end{figure}
}

\newcommand{\figaia}
{\begin{figure}
\includegraphics[width=1.05\linewidth]{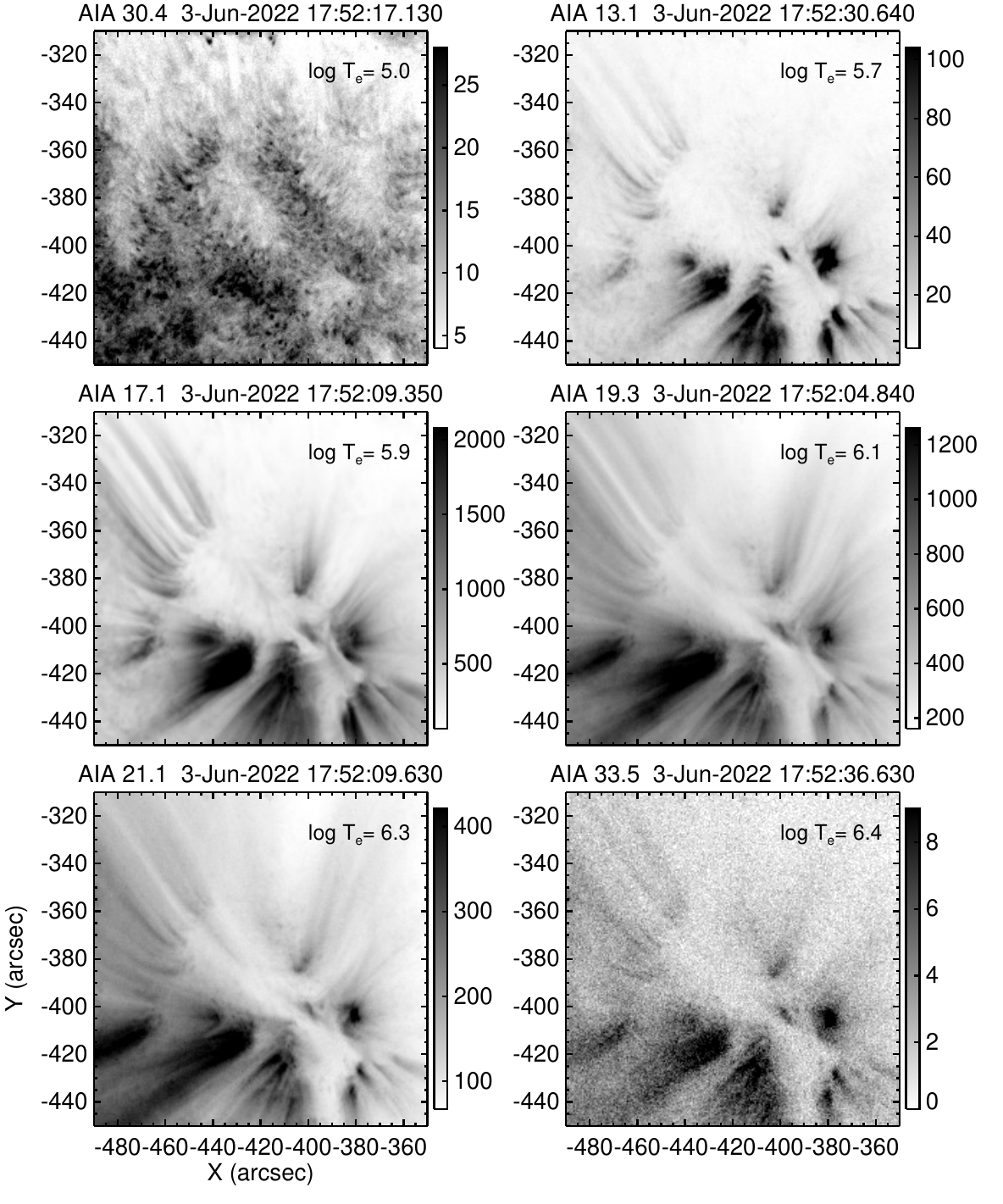} 
\caption{\pja{Images of UV and EUV emission from the AIA instrument 
are shown, obtained close to
the center of the 
ViSP raster scan.
Each image is a sum 
of several images between 17:50 and 17:53 UT. Intensity scales are average data numbers.}}
\label{fig:aia} 
\end{figure}
}

\newcommand{\figwestline}
{\begin{figure}
\includegraphics[width=\linewidth]{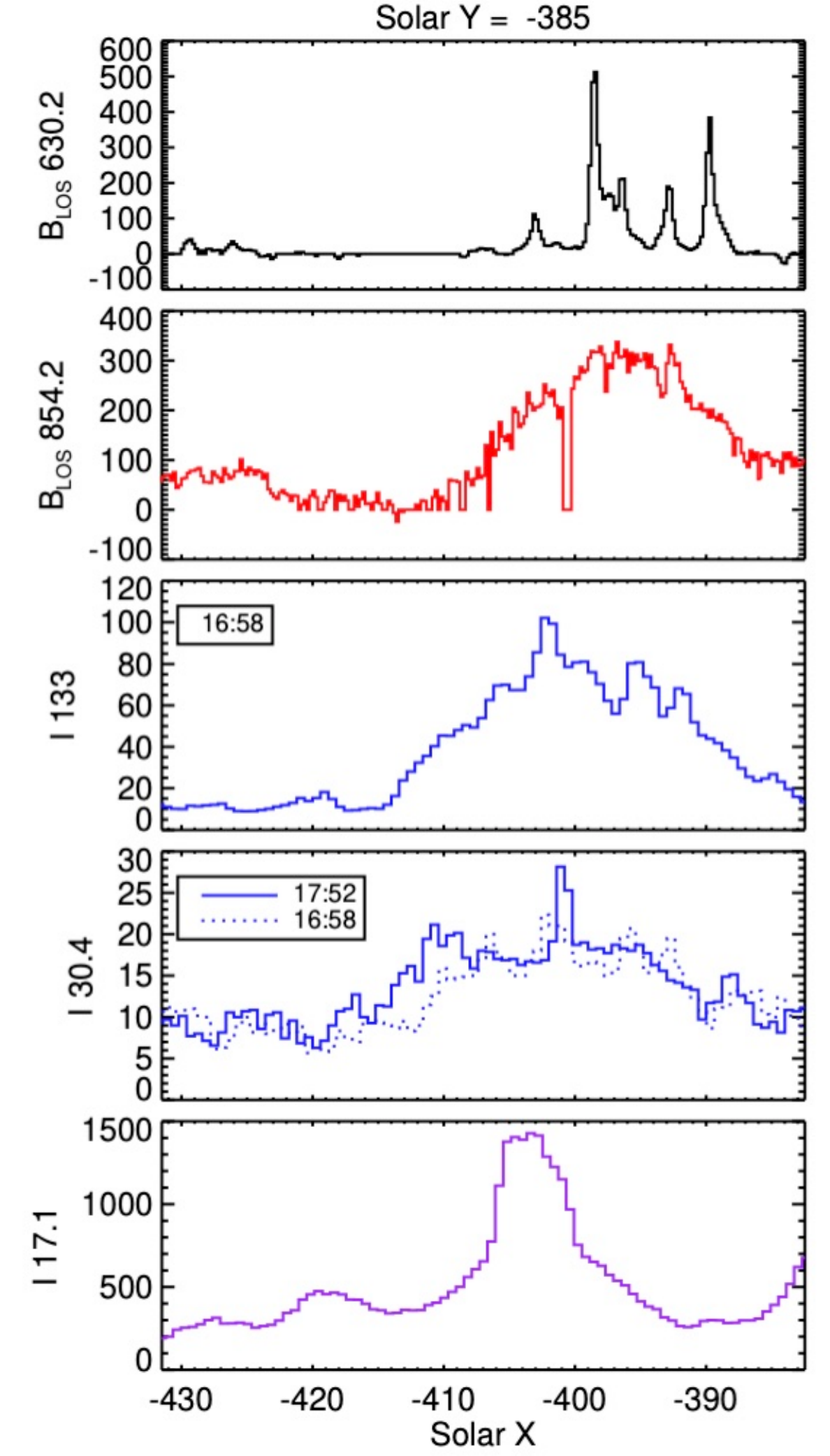} 
\caption{Line plots are shown of several quantities extracted from
the white line of Figure~\ref{fig:westline}.}
\label{fig:westline} 
\end{figure}
}

\begin{document}

\title{\Large{Magnetic fields and plasma heating in the Sun's atmosphere}}


\correspondingauthor{P.  Judge}\affiliation{\hao}

\author[0000-0001-5174-0568]{P.  Judge}\affiliation{\hao}
\author[0000-0002-7791-3241]{L. Kleint}\affiliation{University of Bern, Astronomical Institute, Sidlerstrasse 5, 3012 Bern, Switzerland}

\author[0000-0001-6990-513X]{R. Casini}\affiliation{\hao}

\author[0000-0002-5084-4661]{A.G.~de~Wijn}\affiliation{\hao}


\author[0000-0002-7451-9804]{T.~Schad}\affiliation{\nsomaui}

\author[0000-0003-3147-8026]{A.~Tritschler}\affiliation{\nso}

\date{Accepted . Received ; in original form }


%
%

\begin{abstract}
We use the first publically available data from the Daniel K.  Inouye Solar Telescope (DKIST) to track  magnetic connections from the solar photosphere into the corona.  We scrutinize relationships between chromospheric magnetism and bright chromospheric, transition region and coronal plasmas.  In June 2022, the Visible Spectro-Polarimeter (ViSP) instrument targeted unipolar network within a decaying active region.  ViSP acquired rastered scans with
longitudinal Zeeman sensitivities of 0.25 Mx~cm$^{-2}$ (\ion{Fe}{1}
630.2 nm) and 0.5 Mx~cm$^{-2}$ (\ion{Ca}{2} 854.2 nm).  ViSP was operated in a ``low''
resolution mode (0.214$\arcsec$ slit width, spectral resolution $\mathcal{R}\approx 70,000$) to produce
polarization maps over a common area of 105$\arcsec
\times50\arcsec$. Data from SDO and IRIS are combined to ask:
Why is only a fraction of emerging flux filled with heated plasma? What is the elemental nature of the plasmas?  No correlations were found between heated plasma
and properties of chromospheric magnetic fields derived from the WFA, on scales below
supergranules. Processes hidden 
from our observations control plasma heating. While improved   magnetic measurements are  needed,  
these data indicate that ``the corona is a self-regulating forced system'' \citep{Einaudi+others2021}.  Heating depends on the state of the corona, not simply on boundary conditions.
Heating models based upon identifiable bipolar fields, including 
cool loops, tectonics
and observable magnetic reconnection, are refuted for these regions with unipolar chromospheric magnetic fields. 
\end{abstract}


\section{Introduction}
\label{sec:statement}

In the absence of sufficiently sensitive measurements of magnetic
fields directly beneath the corona, speculations abound concerning the
predominant physical processes responsible for the structure and
heating mechanisms of the overlying plasmas.  Magnetic fields
penetrating from beneath the solar surface have been  established as the
agent responsible for heating the chromosphere and corona, and the
plasma at intermediate temperatures  
\citep{Howard1959,Leighton1959,Reeves1976,Schrijver+Zwaan+Maxson+Noyes1985,Schrijver1987,Schrijver1988,Schrijver+Harvey1989,Cook+Ewing1990,Neupert1998,Fisher+others1998,Schrijver+others1998, Gallagher+others1998,Mandrini+Demoulin+Klimchuk2000, Martinez-Galarce+others2003,Schrijver+Title2003,Doschek+Mariska+Akiyama2004,Loukitcheva+others2009,Parnell+DeMoortel2012,Schmelz+Winebarger2015,Ayres2021,Toriumi+Airapetian2022,Aschwanden+Nhalil2023}.  
Among many others including stellar work, these studies document 
correlations between magnetic fields, plasma emission from chromosphere, transition region and corona.  
While exceptions exist \citep[see, for example the introduction of the paper by ][]{Chitta+others2021}, the bulk of the observational analyses generally find strong correlations.  The 
current paradigm  requires that 
magnetic fields are necessary for significant plasma heating, but that not all
 magnetic fields threading the atmosphere lead to plasma heating.

\pja{These results 
have prompted various 
proposals for magnetic heating mechanisms, such as 
the resonant  
development and dissipation of internal surface waves
\citep[e.g.][]{Ionson1978},
phase mixing 
\citep{Heyvaerts+Priest1983,Mok+Einaudi1990,Howson+others2019} and small scale 
magnetic reconnection 
exhibited in the form of
small flares 
\citep{Parker1988,Parker1994}, driven
perhaps by turbulence 
\citep{Rappazzo+others2008}.
But }
an 
unfortunate aspect of current understanding is that 
multiple solutions to
the coronal heating problem remain actively in contention
\citep[e.g.][]{Aschwanden2001,dePontieu+others2021, Pontin+Priest2022}, despite
decades of coronal
observations and advances in computation  \citep{Judge+Ionson2023}. 
Other elementary 
difficulties still challenge our understanding.  For example, 
although called ``transition
region'' plasmas, the  plasmas responsible for the bulk
of the emission between $10^4$ and $10^6$ K 
may not necessarily 
form a physical transition between warm and hot plasma  
(contrast the perspectives of \citealp{Feldman1983,Dowdy+Rabin+Moore1986,Hansteen+others2014} and \citealp{Judge2021}).  The inability of classical heat
conduction from the corona to account for this ``transition region''
emission below $10^5$ K
\citep[see, e.g.,][]{Cally1990},
including the dominant line of H~L$\alpha$, 
has prompted authors for over
three decades to look for other  explanations
\citep{Mariska1992,
  Judge2021}.  One idea has proven resilient, namely that a host of
unresolved loops lie within supergranular cell ``lanes''
\citep{Dowdy+Rabin+Moore1986}, regions of magnetic fields accumulated
as granular-scale magnetic structures are advected by horizontal
supergranular flows to these lanes. Interaction with the pre-existing
flux leads to bright emission from rapidly evolving small bipolar
loops \citep{Dowdy+Rabin+Moore1986}, and to the development of
embedded current sheets, which when dissipated are also a source for
heating the overlying corona \citep{Priest+others2002}.
\citet{Hansteen+others2014}, using the IRIS instrument
\citep{dePontieu+others2014}, were able to
resolve some cool loops in observations near the solar
limb. However, \citet{Judge2021} argued that cool loops are not abundant enough
to account for transition region emission in the quiet Sun.

\fighe

Our goal is to study \pja{observationally} how
plasma heating is related to 
magnetic structure \pja{measured} in the photosphere and chromosphere obtained using the ViSP instrument
\citep{Dewijn+others2022} at the Daniel K. Inouye Solar Telescope
\citep[DKIST:][]{Rimmele+others2020}.   Novel aspects include not only
chromospheric magnetic fields, but also velocity fields, all 
compared at the highest 
angular resolution and sensitivity with accurately co-aligned heated plasmas from
chromosphere to corona.  We will show that the correlations break down when scrutinized in such detail. 
Since magnetic energy is
converted to heat only on 
small, unresolved scales, we will argue that the generally accepted paradigm for coronal heating may need modifying to be consistent with these new and 
revealing results.

\begin{table*}
\caption{Parameters of ViSP observations on 2022-06-02 and 2022-06-03}
\begin{center}
\begin{tabular}{llrllllll}
      \hline\hline ID & $\lambda$ &mid-scan time & $x_{cen}$ &
      $y_{cen}$ & $dx$ & $dy^\dag$ & rotation & nearest IRIS \\ & & &
      \multicolumn{4}{c}{------------  arc seconds -----------} & clockwise deg. & SJI mid time\\ \hline AODMM &630&
      2-Jun-2022 20:02:23 & -494.7 & -405.0 & 0.2130 & 0.0594&-0.1&
      17:54\\ APJND &854& 20:02:23 & -494.5 & -410.1 & 0.2126 & 0.0388
      &  -0.15& 17:54\\ \\ BRWJV &630& 3-Jun-2022 17:52:23 & -411.1 & -377.0 &
      0.2128 & 0.0594 & -0.18 & 16:58\\ AVORO &854& 17:52:23 & -410.5 & -381.9
      & 0.2124 & 0.0388 & -0.23& 16:58\\ \\ AWOWP &630& 17:23:21 & -416.4 &
      -436.2 & 0.2090 & 0.0594 & -0.15&  16:58\\ AXVLY & 854& 17:23:21 &
      -415.6 & -441.0 & 0.2081 & 0.0387 &-0.20& 16:58\\ \\ BPJDD &630&
      18:21:07 & -311.7 & -377.0 & 0.2120 & 0.0594&-0.10& 20:04\\ BQKZZ
      &854& 18:21:07 & -311.3 & -382.1 & 0.2120 & 0.0389&-0.02&
      20:04\\ \\ BNKVM &630& 18:49:52 & -312.5 & -434.3 & 0.2137 &
      0.0600& -0.10 & 20:04\\ BODXM &854& 18:49:52 & -312.0 & -439.5 & 0.2135
      & 0.0383 &-0.05& 20:04\\ \hline
\end{tabular} 
\end{center}
Pixel sizes $dx,dy$ and rotations 
of the ViSP images are those derived from the alignment with the HMI 
data. The ViSP scans took 26 minutes, IRIS scans were 14 minutes in length.
$^\dag$These values of the 
pixel size along the solar $Y$ direction have been
binned by a factor of two, i.e. they are twice those of the original data.
\label{tab:seq} 
\end{table*}

\section{Observations}
\subsection{VISP data}
\label{subsec:vispdata}

The data were  obtained during a joint campaign with
Parker Solar Probe \citep{Fox+others2016}. 
The region observed and reported here was assigned
the number NOAA 13066 on 5 May 2022. Classified according to the
scheme of \citet{Giovanelli1982}, this is an unusually unipolar
region. More typical ``unipolar'' regions exhibit a ratio of dominant
to subdominant polarity near 9:1, not 50:1 as measured here by ViSP.  AIA
images {\citep{Lemen+others2012} from the SDO spacecraft \citep{Pesnell+others2012} reveal network that is brighter than average, lying at the leading (westward) edge of the plage arising
from the breakup of NOAA 13006.  This work builds on that of
\citet{Judge+Centeno2008} by including measurements of chromospheric
magnetic fields in the \ion{Ca}{2} 854.2 nm line as well as the
photospheric \ion{Fe}{1} 630.2 nm line pair.  These data are
augmented with simultaneous data from instruments on SDO, and
data from the IRIS instrument obtained before and after the DKIST scans.

%
\figfov

The DKIST  targeted  chromospheric network 
\citep{Hale+Ellerman1904}
within the remnant of NOAA 13006,
highlighted in Figures~\ref{fig:he} and \ref{fig:fov}.  The
DKIST experiment ID was eid\_1\_118.  Here we highlight data obtained with
arms 1 and 3 of the ViSP spectrograph, with cameras centered at 603.2 and
854.2 nm (see Table~\ref{tab:seq}), for one scan that began
near 17:39 UT on 3rd June 2022, a period of very good seeing. 
Arm 2 of the spectrograph was  not used. The Table
includes all observations analyzed in our study, and pairs together
data from the two arms, each of which has its own geometry as listed
in the central solar coordinates ($x_{cen},y_{cen}$)  spatial  pixel sizes ($dx,dy$), and rotation angle.  The field of view for arm 1 was
$105\arcsec\times77\arcsec$, while 
for arm 2 it was $105\arcsec\times50\arcsec$ because of  different  magnifications at the image planes.  We present and analyze data from DKIST level 1 datasets labeled 
BRWJV and
AVORO. They represent  data typical of all 
scans, selected
because of a more uniform quality of seeing.
The two wavelength regions contain Zeeman-sensitive spectral
lines of \ion{Fe}{1} and \ion{Ca}{2} formed in the photosphere and
mid-chromosphere respectively.  ViSP was operated in full
spectropolarimetric mode. \pja{It recorded
linear combinations of Stokes parameters 
$I_\lambda,Q_\lambda,U_\lambda,I_\lambda$ with time, parameters which measure intensity $I$, two states of linear polarization ($Q,U$) and circular polarization 
($V$, see \citealp[chapter 1 of][for definitions of the Stokes parameters]{Landi+Landolfi2004}).}

Each scan used the largest available ViSP slit at the spectrograph
entrance, subtending an angle of 0\farcs214, with a slit-limited resolution
of 0\farcs428 ($\approx 300$ km at the Sun's surface), along the solar $X$ direction (E-W). The slit was
oriented in the solar N-S direction, and was moved across the solar image in steps of a full slit-width between integrations.
Below we will find that the resolution measured along the slit (N-S, solar $ Y$ 
direction) is also  $\approx0\farcs4$.  At each slit
position the instrument measured the full state of polarization of the
incoming light, by recording intensities at 10 equally spaced angular
positions of the continuously rotating modulator. The frame rate was
41.357 Hz, and the exposure times were 4.0 ms for both cameras. Thus the camera duty
cycle was $\approx 16.5\%$.  For arm 1 an attenuation filter with an optical density of 0.6 was inserted in the beam.  For each slit position,
the modulation cycle was repeated 12 times (6 full rotations of the
modulator), and each of the 10 modulated signals was co-added over the
12 cycles.  The ViSP performs dual-beam polarimetry to reduce
seeing-induced crosstalk \citep[e.g.,][]{Lites1987}, and records the
two beams of orthogonal polarization on the same detector.
The 
instrument performance model,
verified  
using science verification 
data for the 0\farcs041 slit, 
predicts spectral  resolutions of 76,000 (630.2 nm) and  69,000 (854.2 nm) for the 0\farcs214
slit.   

\subsection{Alignment of different datasets}

At the level of accuracy required below, alignment between various instruments required 
great care.   We adopt
the coordinate
frame defined 
\pja{by the line-of-sight (LOS) field
components from the Stokes $V$ 
measurements }
of the HMI instrument
\citep{Hoeksema+others2018},  against which 
all images were co-aligned.

Line-of-sight magnetograms constructed from the level 1 ViSP data were
aligned 
to better than $\approx0.3\arcsec$, by
eye.   \pja{These alignments
required modest image stretching and 
rotation, as well as 
centering, and were done 
for photospheric lines in 
each ViSP
camera, thus removing any 
image
shifts by 
differential refraction. 
 More precise 
determinations were 
precluded by the fact that
the ViSP is a scanning slit instrument, and 
the Stokes 
$V_\lambda$ signals
evolved
during the 26 minute ViSP scan times. A higher precision
was found to be unnecessary
for our purposes 
\textit{post-facto}, and so
more refined co-alignments
of HMI and ViSP data were not attempted. }

The geometric ViSP
pixel sizes inferred from this alignment (listed in
Table~\ref{tab:seq}) are within 1\% of those in the VISP level 1
headers.  Along the slit (i.e., in the solar $Y$-direction), angular
resolution is set by the telescope system's imaging performance, as
well as seeing.  The nominal pixel sizes along the slit are
0\farcs0294 and 0\farcs0194 for arms 1 and 3 (630.2 and 854.2 nm)
respectively, smaller than the slit width by factors of 7.4 and 11
respectively.  
Given these rates of over-sampling, we
re-binned data 
along the slit direction into pixels 
a factor of two larger before proceeding, 
to handle the many
images more quickly
during processing.
The larger 
bin sizes used,
determined from
the alignment with HMI, are listed in
Table~\ref{tab:seq}.
The alignment quality  can be partly 
assessed using Spearman cross-correlation coefficients, which are listed 
in Table~\ref{tab:spearman}.

\begin{table*}
\caption{Spearman's rank correlation coefficients over small areas (10
  Mm) of BRWJV  and AVORO}

\newcommand{\mb}[1]{\textbf{#1}}  
\begin{tabular}{lrrrrrrrrrrrrr}
    \hline\hline 
  & $B^{854}_{LOS}$ & $B^{854}_{POS}$ & 160 & 170 & 279 & 133 & 140 & 30.4 & 13.1 & 17.1 & 19.5 & 21.1 & 33.5\\
  $B^{630}_{LOS}$      &  0.49&  0.01&\mb{  0.79}&\mb{  0.80}&  0.15&  0.06&  0.06&  0.13&  0.08& -0.13& -0.19& -0.17&  0.08\\
  \hline
$B^{854}_{LOS}$      &      &  0.01&\mb{  0.50}&  0.46&  0.33&  0.06&  0.13&  0.40&  0.06& -0.11& -0.25& -0.26& -0.01\\
$B^{854}_{POS}$      &      &      &  0.06&  0.05&  0.12&  0.07&  0.06&  0.10&  0.06&  0.09&  0.06&  0.09&  0.04\\
160      &      &      &      &\mb{  0.95}&  0.33&  0.22&  0.21&  0.26&  0.16& -0.05& -0.14& -0.09&  0.11\\
170      &      &      &      &      &  0.25&  0.17&  0.17&  0.21&  0.09& -0.12& -0.21& -0.15&  0.07\\
279      &      &      &      &      &      &\mb{  0.63}&\mb{  0.61}&\mb{  0.62}&\mb{  0.57}&\mb{  0.51}&  0.30&  0.42&  0.39\\
133      &      &      &      &      &      &      &\mb{  0.79}&  0.31&  0.44&  0.31&  0.18&  0.31&  0.33\\
140      &      &      &      &      &      &      &      &  0.32&  0.48&  0.37&  0.27&  0.35&  0.36\\
30.4      &      &      &      &      &      &      &      &      &  0.28&  0.12& -0.07& -0.01&  0.17\\
13.1      &      &      &      &      &      &      &      &      &      &\mb{  0.71}&\mb{  0.70}&\mb{  0.68}&\mb{  0.80}\\
17.1      &      &      &      &      &      &      &      &      &      &      &\mb{  0.91}&\mb{  0.94}&\mb{  0.53}\\
19.5      &      &      &      &      &      &      &      &      &      &      &      &\mb{  0.95}&\mb{  0.55}\\
21.1      &      &      &      &      &      &      &      &      &      &      &      &      &\mb{  0.52}\\

   \hline
\end{tabular} 
\label{tab:spearman} 
\newline
\newline
Rank correlation coefficients above 0.5 are in boldface.  The
coefficients in rows and columns for 279, 133 and 140 were computed from contemporaneous IRIS
data obtained near 16:51 UT. 
The other values were computed using data obtained close to 17:52 UT.
Small correlation coefficients of $ 0.33$ and $0.30$ are found
between SDO 30.4 and IRIS 133 and 140 nm 
features obtained  55 minutes apart, \pja{evolving significantly } (Figure~\ref{fig:west}), suggesting that values of $\lesssim 0.3$  indicate  
insignificant  correlation.
\end{table*}

The angular resolution along the ViSP slit was estimated from power and
phase-difference spectra of intensity, which show  coherent power on
all scales up to $\approx 0.15\times$ the Nyquist frequency
($\mathcal{N} = 0.5/dy$). This corresponds to a full width at half maximum of the point spread function---the effective angular resolution---of $\approx0\farcs4$.
The ViSP Feed-Optic telescope that focuses the DKIST incoming beam onto the spectrograph's slit 
was later discovered to be 
slightly out of focus after
this campaign, 
limiting the spatial resolution
achievable.
Somewhat fortuitously, this happened to be close to the sampling-limited
resolution along the orthogonal 
$X$-direction (0\farcs428), thus
the images appear 
consistent with a
symmetric PSF.

Data from SDO HMI
and AIA instruments were acquired with
their nominal cadence throughout the DKIST scan.  
Only the ``45s'' longitudinal magnetic field from HMI 
was used, used only to provide a reference frame 
for all observations.  
Two different methods were used to
align EUV and UV data to the HMI
reference frame. For AIA, we found 
Sun center through optimal correlations 
of the entire field of view of each image
with the 2D Gaussian function
$\exp (-r^2/r^2_\odot)$.  For IRIS,  we then performed
a cross-correlation for the overlapping
field of view of the $160$ nm AIA
images with those of IRIS. These alignments 
are $\pm1$\arcsec{}  or better.
The 
coefficient for 
rank cross-correlation between HMI $B_{LOS}$ and AIA 160 nm listed
in Table~\ref{tab:spearman} is 0.68,
increasing to 0.80 for the 170 nm
band formed in the upper photosphere \citep{VALIII}.

The final column in Table~\ref{tab:seq} shows the start times
of the nearest IRIS scans because no simultaneous measurements were
made with the ViSP scans.  Each IRIS scan ran for 14 minutes.  Of these, 
the two DKIST scans 
AWOWP/AXVLY and BPJDD/BQKVV began within half an hour of the IRIS scans.
Nevertheless, we focus here on BRWJV/AVORO because of the higher quality 
seeing conditions.   This means that we will examine the three IRIS 
data products (133, 140 and 279 nm slitjaw images, 
dominated by emission from lines of \ion{C}{2}, \ion{Si}{4} and \ion{Mg}{2}
respectively) obtained 
an hour before the DKIST scans began.   The correlations reported 
in Table~\ref{tab:spearman} are from simultaneous data from SDO and IRIS,
and from simultaneous data from SDO and ViSP.  Those between ViSP and 
IRIS are \textit{not} simultaneous measurements, the effects 
of this mis-match are discussed below.

It is important to stress that image alignments better than 1\arcsec{}
proved possible only for morphologically
similar images, such as AIA 13.1, 17.1, 19.3 images, HMI and DKIST magnetograms and 
AIA 170 and 160 images, and IRIS slitjaw images at 133, 140 and 279 nm images with
AIA 170 nm.   The accuracy of alignments 
of the EUV AIA data with all other
data are therefore subject to  larger errors \pja{of $\approx \pm 2-3\arcsec$.  These uncertainties 
were estimated 
by comparison between the 
values derived above with a careful manual
alignment using 
particular features  across the solar disk.}
Of our conclusions, none is affected by these 
problems, because
no refinements of the alignments
alter the 
lack of correlations which is the novel finding of this study, \pja{even for
mis-alignments 
 $\gtrsim 5\arcsec$ (compare, e.g.,
variations of $B_{LOS}$ with those of 17.1 emission shown in 
Figures~\ref{fig:eastline}, \ref{fig:westline}).}

\subsection{UV and EUV data}

In our analysis of EUV data from the AIA instrument on SDO, we
averaged ten exposures obtained from 17:51 to 17:53 UT to enhance
signal-to-noise ratios (Figure~\ref{fig:fov}), during which time the
solar images rotated by less than 0\farcs3. 
\pja{Figure~\ref{fig:aia} shows 
images from the
transition region 
to 
the corona.
The morphology 
of coronal emission remains similar for all coronal images, 
the plasmas 
emitting in the same fashion at
the temperatures typically sampled by these data.
}
\figaia
These data were
augmented with slitjaw images from IRIS, which scanned the same
region, but only before and after the DKIST scans.  IRIS was
unfortunately pointed to a different active region during the DKIST
scan with mid-time 17:52:23 UT, but IRIS data are available from 16:50:56.700 to 17:05:37.760
UT, ending 41 minutes before the start of the DKIST scan.
In the figures these IRIS data were obtained between
16:51 and 17:10 UT, but were 
rotated to 17:52 UT for direct comparison with the other images. 
Further IRIS
scans only began after 19:57 UT on 23 June 2023.

\figeast

Figures \ref{fig:east} and \ref{fig:west} show close-up views of
DKIST magnetic and intensity data in comparison with the AIA data,
allowing us to relate transition region emission in the \ion{He}{2}
30.4 nm line to be assessed relative to the overlying corona as well
as underlying magnetic fields. A well-known but weak blend within the
30.4 nm bandpass of AIA includes a line of \ion{Si}{11}, whose emission
should be similar to the AIA images of \ion{Fe}{12} at 17.1 nm shown.
Included in Figures \ref{fig:east} and \ref{fig:west} are slitjaw
images obtained by the IRIS instrument in \ion{C}{2} at 133 nm, 
those of 
\ion{Si}{4} (not shown) are morphologically very similar. 
lines, both formed in the lower transition region.  
\figeastline

\figwest

In these figures, the IRIS images, obtained over 45 minutes before
the DKIST scan, are compared with simultaneous AIA data at 30.4 nm as well as
data obtained during the ViSP scan.  The sequence of 24 slitjaw images acquired by
IRIS between 16:51 and 17:05 reveals time-dependent changes typical of
other solar regions \citep[e.g.,][]{Skogsrud+others2015}.  The 30.4 nm
panels of Figures~\ref{fig:east} and \ref{fig:west} show  that
the morphology of the IRIS \ion{C}{2} images at 133 nm are 
 quite different.  
\ion{Mg}{2} 279 and \ion{Si}{4} 140 nm data (not shown) are  
 morphologically 
 similar
 to those at 133 nm.  
Under standard assumptions of the formation of these EUV lines
\citep{Lang+Mason+McWhirter1990}, these enormous differences are not possible, and
the morphology would be similar on all observed scales.  
The implications of such radical departures from theories of line formation as well as
models for energy transport though the 
``transition region'' 
will be
speculated upon below.
\figwestline

\pja{Variations of  coronal images on June 3rd 2022 are far less dramatic
than those observed in transition region plasmas,
the 1 hour differences being
highlighted in the AIA 17.1 images of Figure~\ref{fig:fov}.
The individual  
plasma loops persist within at least two of three 
images acquired at 17:29, 17:52 and 18:29 UT, slowly coming and going within the same magnetic
structure. This is consistent with a 
slow thermal cooling time of
10-20 minutes \citep[e.g][]{
Kuin+Martens1982}.}

Figures~\ref{fig:eastline} and
\ref{fig:westline} show 
slices through some of the 
images of Figures~\ref{fig:east}
and \ref{fig:west} respectively, reinforcing the 
above statements, and showing
quantitatively the remarkable
differences between 
properties of
photospheric, chromospheric, transition region and coronal plasmas.
\pja{These differences are plainly evident 
even at the angular resolution of the EUV SDO images, which
is at best 1\arcsec, compared
with 0\farcs4 for the ViSP data.}

\subsection{Chromospheric velocity fields from DKIST data}

Using the intensity profiles of the \ion{Ca}{2} line, we derived
line-of-sight velocity shifts $v_{\rm LOS}$ and line widths $w_{\rm LOS}$ in
two ways.  We are mostly concerned with conditions at the base of the
corona, and so we focus on the \ion{Ca}{2} line at 854.2 nm.

Shifts were derived from fits to the absorption line profiles using
Gaussian functions.  These were checked against finding the spatial
average of all profiles and finding the maximum cross-correlation of
each profile with the average, measured in pixels, and using a
parabola to find the peak of the cross-correlation function.  Both
derivations yield values close to the central wavelength of the line cores,
the minimum in intensity.

The non-thermal Doppler width $\xi$, will be used below to estimate
Poynting fluxes entering the corona:
\begin{equation} \label{eq:width}
    \xi =   \sqrt{w_{LOS}^2 - \frac{kT}{A\,m_H} - s^2 }  \ \ \ \mathrm{cm~s^{-1}}.
\end{equation}
Here, $w_{LOS}$ is the Doppler line width (i.e., the measured full width at half-maximum line depth 
divided by $2\sqrt{\ln 2}$), 
$k$ is Boltzmann's constant, $A$ the atomic mass of the 
radiating ion,
$m_H$ is the mass of a hydrogen atom, and we adopt a kinetic
temperature $T$ near 7000 K.  For calcium $A=40$ and 
$\sqrt{kT/Am_H} = 1.2\ 10^5$ cm~s$^{-1}$.
The correction for the
spectrograph PSF is given by $s^2$ where 
\begin{equation}
s =  \frac{1}{2\sqrt{\ln 2}} \,\frac{c}{\mathcal{R} }
= 2.6\ 10^5 \ \ \ \mathrm{cm~s^{-1}
}\end{equation}
which is the $1\sigma$ width 
of the spectrograph PSF with resolution
${\mathcal R}=69,000$.  
For 
pixels with widths $d\lambda =0.00194$ nm,
each pixel
corresponds to 
a Doppler shift of
0.68\ 10$^5$ cm~s$^{-1}$.

Widths $w_{LOS}$ for such a strong line as 854.2 nm, formed 
over many scale heights across the photosphere and chromosphere,
present different
challenges.  Firstly the line is broadened not only by unresolved
plasma motions (kinematics), but also by ``opacity broadening,'' where
photons trapped in the line core are occasionally shifted to wing
wavelengths where they escape.  This is tied to another basic
difficulty in that the ``width'' of such a strong line, formed over
several scale heights in the atmosphere, reflects conditions over the
whole atmosphere from which photons escape to space.  In the absence
of a full ``inversion'', solving for 
atmospheric thermal and magnetic properties, 
we estimated non-thermal line widths in
the upper chromosphere by using wavelength moments of the line depths
(i.e., subtracting the line profiles from the continuum prior to
calculating wavelength-weighted moments).   Sobering systematic 
errors arising from  
inversions have been recently highlighted by 
\citet[][see their section 6]{Centeno+others2023}, even for
photospheric spectral lines. Such problems motivated our choice to use
the far simpler,   less accurate but perhaps
more robust, WFA. 
Tests with Gaussian fits to the core revealed that the derived
values reflect  estimates of unresolved plasma motions in the upper chromosphere
near and above 1400 km above the continuum photosphere  \citep{Cauzzi+others2008}.

\subsection{Magnetic fields from DKIST data}

To derive magnetic fields from ViSP profiles, we used the weak field
approximation (WFA, see \citealp{Landi+Landolfi2004}) for both the
pair of \ion{Fe}{1} lines (630.15, 630.25 nm) as well as the
\ion{Ca}{2} 854.2 nm line.  A photospheric line of \ion{Si}{1} at
853.8015 nm within the 854.2 nm frames was also treated using the WFA,
but only to confirm the close optical alignment between the 630 and
854 nm raster images obtained by ViSP.  In the WFA we have, for
magnetic fields along the line-of-sight $B_{LOS}$
\citep{Landi+Landolfi2004,Centeno2018},
\begin{equation}
V_\lambda = -\Delta\lambda_B\, \bar{g} \cos\theta\, \frac{d I}{d
  \lambda}\label{eq:wfablos}
\end{equation} 
where the Zeeman broadening parameter is
\begin{equation}
   \Delta\lambda_B = 4.6686\times10^{-11}B\, \lambda^2,
\end{equation}
with $\lambda$ in nm, and $\bar{g}$ is the effective Land\'e factor
for each line (1.67, 2.5 and 1.1 for 630.15, 630.225 and 854.2 nm
respectively).  $B_{LOS} = B\cos\theta$ was derived in several ways, the plots showing
data from the ratio of integrals of $V_\lambda$ and $dI/d\lambda$ over
$\lambda$.  However, following \citet{Kleint2017}, we also made a least-squares fit of $V_\lambda$ to the numerical derivative of $I$ to derive
LOS magnetic flux densities to estimate uncertainties.  RMS variations
of the fits were found to be 1.2 and 8 Mx~cm$^{-2}$ for the 630.2 and
854.2 line respectively. These are larger than the $1\sigma$ random
variations estimated using power spectra of spatial variations derived
in Table~\ref{tab:sens} (see Figure~\ref{fig:sens}).

For the magnetic field component in the plane of the sky (POS), and
away from the line center wavelength $\lambda_0$, we have
\begin{equation}
P_\lambda = \sqrt{Q_\lambda^2 +U_\lambda^2} = \frac{3}{4}\,
\Delta\lambda_{\rm B}^2\, \bar G \sin^2\theta\, \displaystyle
\frac{1}{\lambda - \lambda_0}\displaystyle \frac{d
  I}{d\lambda} \label{eq:wfaQwing}
\end{equation}
\noindent where $\Delta \lambda_B$ is defined by $B_{POS}$ instead of
$B_{LOS}$, and $\bar G$ is the Land\'e factor for $B_{POS}$
\citep{Landi+Landolfi2004}, whose values for 630.15, 630.25 and 854.2
are are 3.29, 6.18 and 2.25 respectively.  Again a least-squares fit
of $P_\lambda$ to the right-hand-side of equation~(\ref{eq:wfaQwing})
was used to derive $B_{POS}$.  The azimuth of the magnetic field in
the POS, $\chi$, is given by
\begin{equation}
\frac{U_\lambda}{Q_\lambda} = \tan 2\chi
\label{eq:azimuthwfa}
\end{equation}
with an ambiguity of 180$^\circ$.

Subject to noise and sensitivity levels, these ViSP data probe the
magnetic roots of the corona with a resolution in the solar $X$
direction of $\approx 310$ km.  For simplicity, we will examine data
as if it were acquired with square pixels of $0\farcs214$ on each
side.  For each exposure, pixels along the slit of size 0\farcs0294
and 0\farcs0194 oversample these square pixels by the factors of
$f_y=$ 7.4 and 11 given above. 
After taking into account the modulation efficiency of the ViSP and the configuration-dependent beam imbalance (mainly caused by the grating polarization), the estimated polarimetric sensitivities are
$\approx 1.2\times 10^{-3}$ measured in the continuum. 
Binning by 
$f_y=$7 pixels along the slit, we estimate a sensitivity of $\approx  4\times 10^{-4}$.  This
level is comparable to the best achieved
with other grating spectropolarimeters \citep{delaCruz+Navarro2011},
smaller by a factor of roughly five than recent observations from an imaging
spectropolarimeter \citep{EstabanPozuelo+others2023}. The best
polarization sensitivity is perhaps $10^{-5}$ obtained by the
far-lower resolution ZIMPOL instrument \citep{Povel1995}.

A polarization sensitivity of $4\times10^{-4}$ per pixel at each spatial pixel translates to $1\sigma$
flux densities from equation~(\ref{eq:wfablos}) of 0.25 and 0.5
Mx~cm$^{-2}$ for the 630.2 and 854.2 lines respectively. These
sensitivities were not, in practice, achieved (see Table~\ref{tab:sens}). While the 
least-squares fits 
effectively integrate over several wavelength
pixels, owing to non-ideal effects such as residual calibration
errors, image motion, atmospheric perturbations and crosstalk, the
measured sensitivities are
larger. 

The
sensitivities of Table~\ref{tab:sens} were estimated by integrating
white noise from spatial power spectra of pixels along the ViSP slit
with the smallest signals (Figure~\ref{fig:sens}).  These include
co-addition of multiple pixels along the slit and dispersion
directions, roughly 25 co-added pixels for the 854.2 nm
line. Therefore they should be compared with ideal values smaller by a
factor of 5 than given above. We conclude that the ViSP instrument
exhibited noise levels several times those estimated under ideal
conditions.

\begin{table}
\caption{Measured $1\sigma$ sensitivities to magnetic fields }
 \begin{center}
\begin{tabular}{llllll}
      \hline\hline Inst. & Line & $B_{LOS}$ &Flux/ & $B_{POS}$ &
      Flux/\\ & & & elem. & &elem. \\
\hline HMI &\ion{Fe}{1} 617.3 & 2.7 & 3.6 &\ldots & \ldots\\ ViSP
&\ion{Fe}{1} 630.2 & 0.7 & 0.15 & 2.3 & 0.5\\ ViSP &\ion{Ca}{2} 854.2
& 2.5 & 0.6& 44 & 10\\ \hline
\end{tabular} 
      \end{center}
\label{tab:sens} 
Flux densities are in Mx~cm$^{-2}$, fluxes per resolution element
(Flux/elem.) are given in units of $10^{15}$ Maxwells.  A ``pixel
length'' of 0.2\arcsec{} along the ViSP slit was adopted, multiplied
by the slit width to find the approximate area of the effective
resolution elements.
\end{table}
\figsense

The total unsigned flux from the WFA applied to the photosphere under
the common area of overlap is
\begin{equation} \label{eq:phi}
    \Phi = a\ \sum_{pixels} \sqrt{B_{LOS}^2+B_{POS}^2}
\end{equation}
with $a$ the area of one pixel.  Within the uncertainties the WFA
fluxes given by equation~(\ref{eq:phi}) are equal in both the 630.2
and 854.2 lines, both yielding a flux of $6.0\times10^{21}$ Mx.  The line-of-sight flux of HMI is 
 $1.7\times10^{21}$ Mx compared with $2.0\times10^{21}$ Mx for the ViSP
630.2 line.  Given the level of accuracy of the WFA and the factor of
at least two difference in angular resolution, this difference is
neither surprising nor significant.

Vector magnetic fields from the WFA are shown in
Figure~\ref{fig:vfield}, along with the rms speed $\xi$ from
equation~(\ref{eq:width}).  When multiplied by $B^2$ this leads to a
quantity which is a crude upper limit on Poynting flux up into the
corona (see section \ref{subsec:connect}, shown in the lower left
panel), in units of erg~cm$^{-2}$~s$^{-1}$.  Clearly \textit{this
  quantity from this particular region has little spatial correlation
  with the coronal emission}, or with emissions from plasma observed
by IRIS or SDO.

The WFA
signals in $B_{LOS}$ 
in Figures \ref{fig:east} and \ref{fig:west} include
detection of the 
fields in the chromospheric 
``canopy'' \citep{Giovanelli+Jones1982}. 
This is readily seen in the  signals 
on either side of the ridges of
intense chromospheric magnetic
fields (the $B_{LOS}$ panels of
Figures \ref{fig:east} and \ref{fig:west}).   The diffuse fields
to the north and west of 
the intense magnetic network
are closer to Sun center 
than those to the south and east,
and of opposite sign. This is
exactly as expected as the fields
become more horizontal farther from the network.  Thus it seems
that the coronal base has unipolar field
extending across much larger areas than just the areas defined, say, by the more intense
magnetic fields observed 
in the photosphere and chromosphere.

We also analyzed the other ViSP scans along with SDO and IRIS data
obtained during the same campaign (Table~\ref{tab:seq}). These all
sampled mostly unipolar network regions associated with NOAA 13066,
and were selected from more scans based upon the quality of ViSP 
quick-look images of granulation.  The findings
reported here for the BRWJV and AVORO data appear to be typical of all these
datasets.

\figvfield

\section{Analysis}

\subsection{Context and summary of observational results}

By themselves, the DKIST data offer little
beyond what is known about chromospheric magnetic fields in relation
to the lower atmosphere.  
For example, a recent similar study examined much quieter areas
  of the Sun's disk
\citep{EstabanPozuelo+others2023},  focusing on chromospheric magnetic fields alone.   They found that, even
in quiet regions
of significant mixed polarity, larger areas of strong network are stable over at
least 18 minutes,
exhibiting little structure below 0.5$\arcsec$ (refer to their Figure 10). In another relevant article,  \citet{Chitta+others2017}
related high resolution photospheric magnetic fields from 
the IMaX experiment 
\citep{Martinez-Pillet+others2011} on
the second SUNRISE balloon
mission \citep{Solanki+others2017}, 
to coronal loops, in a region of
emerging magnetic flux.  They concluded that 
\begin{quote} 
``mixed polarity fields can be found at the base of coronal loops where magnetic flux cancellation events are possible''
\end{quote}
and surmised that 
such bipolar regions may lie at the root of all coronal loops
that were otherwise observed to be unipolar, at lower
spatial resolutions.

The present paper differs in several ways. First it
fills in the enormous gap between photospheric and coronal conditions
by sampling plasmas
across the entire atmosphere. Secondly 
we do not actively seek relationships between observed macroscopic  features, 
\citet{Chitta+others2017} for example point to relations between  chromospheric ``anemones'' and 
 ``X-ray jets'' \citep{Shibata+others2007}.  \citet{Chitta+others2017} 
related both of these phenomena to
 bipolarity in the underlying magnetic fields. 
 This difference in approach  
 may appear
 subtle, but our approach may help avoid 
 the insidious issue of
 confirmation bias
\citep{BarkerBausell2021}.  For example, along the dashed blue line highlighted in Figure~2 of
\citet{Chitta+others2017}, we see many  coronal
plasma loop footpoints which are not associated with bipolar features, and 
to some extent, \textit{vice versa}. 
 
On all scales,  Figures \ref{fig:east} and \ref{fig:west} are rich in
information. Both correlations and 
the lack of correlations are equally significant
and important. 
The lack of correlations 
is a result that is robust against the relatively low accuracy of
co-alignment 
between EUV 
and ground-based images 
(2-3\arcsec).  Viewed 
within the current paradigm for coronal heating  (section~\ref{sec:statement}), these images open up a number of questions concerning the elementary
structure of the solar atmosphere and corona in relation to  magnetic fields emerging from  beneath.  There
is no clear conflict between earlier work and our observational results. Instead 
our work brings together data from across 
the entire atmosphere, relating all
evidence of plasma heating to properties
of chromospheric magnetic fields and 
Poynting flux proxies. 

Particularly striking in our study is the lack of 
correlation
between magnetic fields, estimates of Poynting flux 
and overlying bright plasmas, when scrutinized
on scales below the 
chromospheric network 
(1-20 Mm). 
Table~\ref{tab:spearman} lists rank correlation
coefficients from within the  $12\arcsec\times12\arcsec$ 
area centered near the 
coronal footpoint 
seen in Figure~\ref{fig:west}.
There is little correlation between
variables when the coefficient is less than about 0.3 (as assessed using 
non-contemporaneous data, see
the text
in the Table).  When computed over the $105\arcsec\times50\arcsec$
field of view, correlations between UV and EUV increase, revealing, as they should, the
well-established correlations on large scales.  But it is at and below
these small
scales that irreversible heating occurs, as
expected from theory \citep{Kuperus+Ionson+Spicer1981, Judge+Ionson2023}.  Also, the utter lack of
small-scale correlation of the EUV and UV data with photospheric magnetic fields
is a sobering reminder of the difficulties faced when using 
photospheric magnetic fields alone to infer 
coronal properties, as 
has been common since the late 1960s 
\citep[e.g.][]{Altschuler+Newkirk1969,
Bale+others2019}, and  continued using far 
higher quality data 
by \citet{Chitta+others2017}, among many others.

We can summarize more particular results which can be judged by inspection of our figures:
\begin{itemize}
    \item Chromospheric vector magnetic fields are detected by the
      ViSP,  not only over 
      unipolar photospheric fields  but also extending outwards across the interiors of network cells 
      as a magnetic ``canopy'' \citep{Giovanelli+Jones1982}.
      \item These extensions reach across and over areas of photospheric fields of opposite polarity  lying beneath, consistent with expansion and
      volume-filling of the unipolar magnetic field in the upper chromosphere.
      \item Magnetic flux densities, and estimates of upper limits for
      Poynting fluxes, bear no obvious correlations with plasma
      radiating from the atmosphere above.
    \item Statistical  correlations previously studied in lower resolution
      observations of photospheric magnetic fields and the overlying
      atmosphere begin to break down below scales of about 10 Mm (section~\ref{sec:statement}). 
\end{itemize}
Specific to images of the intensity of transition region 
lines, we find 
\begin{itemize}
    \item Bright emission exists only within strong unipolar
      chromospheric magnetic fields, and
    \item it is far weaker over the
      opposite polarity fields surrounding unipolar
          network boundaries in the photosphere.  
      \item IRIS images in spectral lines of ions of \ion{C}{2} (and
      \ion{Mg}{2} and \ion{Si}{4}) differ radically from those of \ion{He}{2}
      images from AIA, obtained simultaneously (Figures~\ref{fig:east} and 
      \ref{fig:west}).
    \item Bright coronal emission arises from footpoints with almost
      no correlation with the IRIS and \ion{He}{2} AIA images.
\end{itemize}

The last two points are most likely related to  the documented failure of standard emission measure
analysis to account for Li- and Na-like
\citep{Burton+others1971,Dupree1972,Judge+others1995}, and 
helium lines
\citep{Jordan1975,Jordan1980a,Andretta+Jones1997,Macpherson+Jordan1999,Smith+Jordan2002,Smith2003,
Judge+Pietarila2004}. As noted by 
\citet{Judge+Ionson2023},
transition region plasmas
contain only a fraction of the mass 
even of the corona, so that
their brightness is readily influenced by
changes of mass, momentum and energy from
both chromosphere and corona. It 
may be that the number of
``discrepant'' lines 
is large enough to question
the general validity of underlying 
assumptions \citep{Lang+Mason+McWhirter1990},
particularly of the statistical equilibrium 
of plasma with electrons described by
thermal distribution functions.

\subsection{Broader implications}

The radical differences measured between vector magnetic fields in the photosphere (broken-up, fractal-like structures) and chromosphere (smooth structures) imply that 
the chromosphere takes a primary role in
determining the nature of the magnetic field and hence its free energy
entering the corona. 
Qualitatively similar findings are readily
seen in previously published work 
studying chromospheric 
magnetism (see, for example,
Figure 1 of 
\citealp{Socas-Navarro2005a}
or Figures~5 and 13 of \citealp{Anan+others2021}).  Thus, 
while
not a new observational result, the main thrust of the present paper is to explore and emphasize 
previously unacknowledged
consequences of 
this basic fact.   We can also conclude that earlier
work which has not included 
measurements of chromospheric
magnetic fields, such as
described in extensive reviews
by \citet{Carlsson+others2019} and \citet{dePontieu+others2021},
has tended to focus upon thermal chromospheric fine structure and dynamics.  
This has
diverted attention
away from 
the coronal heating 
problem, which primarily  involves magnetic free energy, not the 
energy of ordered flows in the chromosphere \textit{per se}.

Images of  $B_{LOS}$, derived from the chromospheric \ion{Ca}{2} 854.2 nm line, 
are spatially 
smooth. They 
depart strongly with those of the line's
core, velocity and width images
(Figures 
~\ref{fig:east}, \ref{fig:west} and \ref{fig:vfield}), forming 
in the upper-middle chromosphere
(\citealp{Cauzzi+others2008}).
The thermal fine structure is, in terms of force and energy balance,  mostly a thermal
``ornament'' on a far less structured 
magnetic field which contains the bulk of
the stress and energy density.   
Consequently, in the absence of routine measurements of  
magnetic fields threading through the chromosphere, it is not surprising that 
little progress has been made in identifying heating mechanisms
of overlying, hotter plasmas.

The magnetic field measured 
from the WFA in the 
core of the \ion{Ca}{2} 854.2 nm (formed near 1400 km in 
statically stratified models)
is predominantly of a single polarity.  The modest
amount of opposite  polarity photospheric magnetic flux 
has no signature in the chromospheric measurements.
This novel result seems to deny 
models of overlying plasma heating based on multipolar fields over the observed regions.  Such models include ``cool loops'' \citep{Dowdy+Rabin+Moore1986,Antiochos+Noci1986,Hansteen+others2014}, ``tectonics'' \citep{Priest+others2002}, and 
suggest that magnetic reconnection, if important,
must occur only through tangentially
discontinuous components of the vector field,  with amplitudes far smaller
than the strong guide field.

We must stress that our results do not apply
to regions of active flux emergence or of mixed-polarity quiet Sun, 
but to the continued heating of long-lived, mostly unipolar network structures.  Nevertheless 
here we have been able to
refute an
entire class of models for emission
from the network and associated 
coronal loops. This kind of refutation
is badly needed to make progress in
plasma heating problems \citep{Judge+Ionson2023} given the current proliferation of ideas 
which have remained poorly challenged
by critical observations. 

Lastly, the strong correlations
between coronal emission 
and  photospheric magnetic 
field measurements stressed in the literature do not extend down to scales below a few Mm.  Again
this is not a new result, 
the question of why only 
certain bundles of flux 
contain brightly emitting plasma
has been of concern for many 
years \citep{Fort+Martres1974,Bray+Loughhead+Durrant1984,Litwin+Rosner1993,Gurman1993}.  As recognized
by these authors and 
re-emphasized by \citet{Judge+Ionson2023} and here, the implications regarding 
heating mechanisms are important
and are discussed next. 

\subsection{Physical connections from chromosphere 
to corona}
\label{subsec:connect}

\figbrwjv
Ideally, as a first 
step to connect 
chromosphere and corona, 
we would measure the 
 Poynting flux vector
\begin{equation}
\frac{1}{4\pi}\vec{(u \times B) \times B}   \equiv 
\frac{1}{4\pi} 
\left\{ {(\vec B} \cdot {\vec u}) 
{\vec B}
- B^2 {\vec u}
\right\}
\label{eq:poynting}
\end{equation}
to determine the energy flux directed upwards into the overlying
plasmas.  Below it will be convenient to write 

\begin{equation}
\vec{B}(\vec{r},t) = \vec{B}_0(\vec{r}) + \vec{b}(\vec{r},t) \label{eq:decomp}
\end{equation}
where $\vec{B}_0(\vec{r})$ represents the large-scale observable field, averaged over time, and $\vec{b}(\vec{r},t)$ 
is the unobservable time-variable field
associated with 
irreversible magnetic energy dissipation.  Within the corona $\vec{B}_0$
may be close to a potential or force-free
field, for example. 
Averaged over many dynamical times, 
$\langle \vec{b}\rangle = 0$,
and 
$\langle \vec{B}\rangle =  \vec{B}_0$.

The radiation
losses from the chromosphere exceed those
from the corona by at 
least an order of magnitude, 
in a real sense the corona is
formed from the energy flux left over from heating the chromosphere.
Therefore
the energy flux 
from equation~(\ref{eq:poynting}) should ideally be evaluated as close to the coronal base, i.e. the top  of the chromosphere, as possible.  
\citet{Athay+White1978}
evaluated not the 
Poynting flux but the 
mechanical (acoustic) flux 
near the top of the chromosphere, showing
that acoustic heating 
was insufficient to 
account for coronal heating in general. 

In using
equation~(\ref{eq:poynting}), we must note that 
Zeeman measurements 
of $\vec{B}$ do 
not determine 
the sign of the field transverse to the line-of-sight.  The second term of equation~(\ref{eq:poynting}) is clearly
unchanged under sign changes of
any components of $\vec{B}$, but
the first term is not.
Therefore we cannot, even in principle, use the Zeeman effect to determine Poynting fluxes unless this transverse field ambiguity can be 
resolved by some means not
contained in the Zeeman 
data themselves.  

Worse still, $\vec{u}$ is also not determined through observations, instead we merely have the line-of-sight velocity shift $u_\parallel$ and the line width
$\xi$.  
The nearest measurable quantities to the  Poynting flux
vector are just the scalar quantities 
\begin{eqnarray} \label{eq:scalarone}
    &v& B^2/8\pi\ \ \mathrm{and} \\
    &\rho& \xi^2 c_A,   \ \ \ 
 \label{eq:scalartwo}
\end{eqnarray}
where $\xi$ is interpreted as
the amplitude of Alfv\'en waves, and 
$c_A = B / \sqrt{4\pi\rho}$
is the group speed of the waves. 
It is easy to see the equivalence of equations~(\ref{eq:poynting}) and 
(\ref{eq:scalartwo}) by noting that
$\frac{1}{2}\rho \xi^2 = b^2/8\pi$ in an Alfv\'en wave,
where $b$ is the magnitude of
the magnetic oscillation, propagating with group speed $c_A$.  Each of these crude ``proxy'' quantities
clearly exceeds the magnitude of the actual Poynting flux
available for heating, owing
to the inequalities 
associated with the double vector
product of equation~(\ref{eq:poynting}), and the fact that 
the sign of the upward 
propagating flux is not known
from the observables derived here.   Nevertheless in 
Figure \ref{fig:brwjv} we
compare images of $\xi B^2$ and $\rho \xi^2 c_A$, as determined using the \ion{Ca}{2} 854.2 nm line, with AIA 30.4 and
17.1 nm images.  These images are gross over-estimates 
of the actual Poynting flux into
the overlying corona from
the chromosphere.  The comparison is meaningful only in the limited  and weak sense 
that if the actual Poynting 
fluxes are some fraction 
of the data plotted, and 
in the upward direction, 
then
we should see correlations between coronal emission and these quantities.   In comparing 
with coronal images there is no
clear correlation. 

These considerations might be pushed 
further by examining the individual images of the 
components 
of $u$, $\xi$, 
$B_{LOS}$ and $B_{POS}$
(Figure~\ref{fig:vfield}), which 
have their own particular spatial 
patterns.     The 
(albeit noisy) components of
vector magnetic fields over the 
magnetic network
vary by less than a factor of 2 (Figure~\ref{fig:vfield}, 854.2 panels), loosely defining network as areas with $B_{LOS}$ measured from  \ion{Ca}{2}
 above about 100 Mx~cm$^{-2}$. 
 Yet over these same regions 
the intensities of coronal emission vary by a factor of 6
or more (bottom panel of Figure~\ref{fig:westline}). 
This suggests that whatever is
heating the overlying corona, 
its signature is either not present in
the data analyzed here, 
or 
that it lies in variations 
in the 
factor  $\vec{\hat{B}}\cdot\vec{\hat{u}}$
in equation~(\ref{eq:poynting}).  
This second case would seem to require peculiarly 
systematic changes in unit vector $\vec{\hat u}$, because observed variations across the observed 
area in 
$u_\parallel$, $\xi$  and 
$\vec{B}$ are too small.
Therefore we suggest that \textit{there are
processes outside of our current ability to measure, which
determine the rate of coronal heating.}  This point of view breaks from the common practice
of seeking signatures in
observable coronal
phenomena \citep[e.g.][]{Peter+others2022}.

Of several  possible explanations, we might suggest 
(1) that the 
heating at EUV wavelengths 
is driven by conditions in the other footpoint, (2) that 
coronal heating occurs only beyond a certain (unknown) intrinsically
critical level of heating 
 \citep{Litwin+Rosner1993}, and
(3) that coronal heating cannot be treated merely as a consequence of lower boundary conditions, instead the heating 
is self-regulating, determining 
itself which bundles of flux are
selected for enhanced heating
\citep{Einaudi+others2021}. 
Case (1) seems  statistically untenable because  all the scans 
in Table~\ref{tab:seq}
qualitatively show the same result (see, e.g., 
Figure~\ref{fig:east}), and 
we should have found at least some 
tighter correlations. 
If case (2) operates then 
the critical level does not
correspond to anything magnetic on observable scales, implying either 
a dramatic
role for unobserved (small-scale) variations 
in $\vec{\hat u}$, or for 
case (3) which has some 
support from numerical turbulence calculations 
\citep{Einaudi+others2021}.   In this sense, 
the non-linearity implied by 
the critical level of heating 
suggested by \citet{Litwin+Rosner1993}
is perhaps just another, less general
case of case (3).

The scalar proxy $\xi B^2 \approx 10^{10}$ 
erg~cm$^{-2}$~s$^{-1}$
vastly 
exceeds the coronal energy requirements of $\lesssim 10^7$
erg~cm$^{-2}$~s$^{-1}$
\citep{Withbroe+Noyes1977}.
This is readily understood, 
 recalling that 
only the non-potential part of
the magnetic field $\vec{B}$ contains energy that is free to provide heating. 
If, as implied in writing 
equation~(\ref{eq:decomp}), 
$\vec{B}_0$
is merely a passive 
player in plasma heating \citep{Sturrock1999},
then \textit{a change in 
approach to the coronal heating problem is warranted,} 
since the conversion of free magnetic energy into heat 
involves only $\vec{b}(\vec{r},t)$, which is not 
directly observable.   The only role
for $\vec{B}_0$ is to fix large-scale topology and $c_A$.  
If strong plasma heating
depends critically
on  variables  outside    observational capabilities, it would then 
explain why
correlations of 
$\vec{B}_0$ and derived 
quantities have led to inconclusive 
results 
\citep{Fisher+others1998,Mandrini+Demoulin+Klimchuk2000,Aschwanden2001}. 
Owing to 
fundamental challenges discussed in chapters~3 and 4 of \citet{Judge+Ionson2023}, there is no guarantee that such ``hidden'' variables 
will be revealed through more advanced 
observations. But suggestions for progress are given in
chapter 5 in the same volume.

In 
the other proxy of Poynting flux 
$\frac{1}{2} \rho
\xi^2 c_A$, $\vec{B}_0$ enters
only as the factor $c_A$,  and the free energy associated with $b$ can be determined through the line widths and mass density.  
$\xi$ is
directly measured, $\rho$ can
be estimated from models, and $c_A$ then estimated from the mesaured $\vec{B}_0$.  
Using 
$\rho=3\times10^{-12}$ g~cm$^{-3}$ appropriate for
the line core  \citep{Cauzzi+others2008}, we find the more interesting energy flux density  of
$\approx 10^8$ erg~cm$^{-2}$~s$^{-1}$, a factor of ten above typical
active-region energy losses.
This number seems encouraging,
suggesting that, whatever the
nature of dissipation, 
energy estimates from
observations and models
might be not unrealistic.
 
We refute cool-loop models
\citep{Dowdy+Rabin+Moore1986, Antiochos+Noci1986} and any models
relying on reconnection with opposite polarity fields (e.g.,
\citealp{Priest+others2002}) to provide heating to power the hot
plasmas above.  
This is basically because 
the 
chromospheric magnetic flux 
beneath the bright EUV emission is unipolar. 
In an appendix, 
we show that invoking small-scale, unobserved 
flux of opposite sign to 
support cool loop models
entails a logical  \textit{reductio ad
  absurdum}.  The absence of correlations between plasma emission
from $\approx 2\times10^4$, to several times $10^6$ K suggests that  the
\textit{observed} transition region plasma is not a thermal interface region
associated with a relatively simple, locally unipolar field \citep[as calculated explicitly in 2D by][for example]{Gabriel1976}.
This argument echos conclusions derived by Feldman and colleagues
based upon data with no magnetic field measurements or the highest 
resolution EUV data
\citep{Feldman1983,Feldman1987,Feldman+Laming1993,Feldman1998,Feldman+Dammasch+Wilhelm2001}.
But our new measurements of
chromospheric magnetic fields enable us to limit further the nature of
plasma heating within the solar atmosphere.  We conclude that the observed transition region plasmas
are heated within locally unipolar magnetic structure by free energy
associated with small-scale unresolved motions and/or non-potential
magnetic fields.  Intermittent current sheets generated by  Parker's fundamental theorem of magnetostatics  
\citep{Parker1988,Parker1994}, by  MHD turbulence  \citep{Einaudi+others2021}, 
or by internal surface waves
\citep{Ionson1978} or phase mixed Alfv\'en waves in disordered plasmas
\citep[e.g.,][]{Howson+others2020} might fit the bill.  Internal
velocity shears generated 
by such 
dynamics might 
cause viscous ion heating \citep{Hollweg1985,Judge+Ionson2023}.
But our essential result is that cool loops cannot account for 
transition region emission revealed by
data obtained down to 0.214\arcsec pixels.

The intensities of lines of helium ions, formed
in transition region plasmas, 
are anomalous
\citep[e.g.][]{Jordan1975,Jordan1980a,Andretta+Jones1997,Macpherson+Jordan1999,Smith+Jordan2002,Smith2003,
Judge+Pietarila2004}.
They are especially sensitive to such
non-equilibrium physical processes \citep[apparently diverse 
proposed processes were
unified in the work of][]{Pietarila+Judge2004}.  The
spatial and temporal differences between all the IRIS data and
the \ion{He}{2} AIA behavior, together with the well documented 
excess emission from 
helium lines 
suggest that helium emission is sensitive
to high energy tails in particle and/or photon distribution functions,
or perhaps long recombination times \citep{Jordan1980a}.  The helium
spectra, presenting a long-standing spectroscopic problem within
transition region plasma, may ultimately shed more light on heating
mechanisms in locally unipolar magnetic fields.
It has been long recognized that the EUV helium lines
are particularly large contributors to
the formation of the ionosphere and its
variability 
\citep{Woolley+Allen1948}, yet the basic 
formation mechanisms 
within areas of active Sun remain 
poorly understood.

\subsection{A curious coronal footpoint}

In Figure~\ref{fig:west}
we noted a curious 
correspondence between the coronal footpoint centered at  $X=-403,Y=-385$, and ring-like structures in the core of the 854.2 chromospheric line and 30.4 nm emission line from 
at 17:52:17 UT.   The 
alignment of these images 
is accurate to $\approx2-3\arcsec$ or better.   No revised alignments within such uncertainties  lead to
better spatial correlations. 
However, if we were to assume that the nominal
alignment is significant,
then we can note
interesting properties:
\begin{enumerate}
    \item The underlying photospheric line of sight magnetic field shows no
    morphological resemblance
    to the overying radiating
    plasmas.
    \item The chromospheric magnetic field directly underlying the structure (Figure~\ref{fig:west})
    is of a uniform strength, 
    lower than some of the neighboring fields.  Hints of narrow opposite polarity fields
    (seen at $X=-404$ to
$-402$, $Y=-386$ in Figure~\ref{fig:west})
are found to be artifacts from multi-peaked intensity profiles (see below). 
    \item  Estimates of the magnitude of the Poynting flux (Figure~\ref{fig:brwjv}) 
    have no correlation with this feature.
\end{enumerate}
The nature of 
the apparent structure 
is also guided by  
Stokes profiles \pja{$I_\lambda,Q_\lambda,U_\lambda,V_\lambda$
of the 854.2 nm measurements. }
The intensity ($I_\lambda)$ profiles of
854.2 along the brighter ring
all show extra emission in the core,  leading to 
self-reversed profiles 
familiarly observed in
the \ion{Ca}{2} H and K lines 
\citep{Linsky+Avrett1970}.  The
854.2 nm $V$ profiles across this area are consistent
with the corresponding WFA (equation~\ref{eq:wfablos}). 
Those hints of opposite
polarity noted above  
are not physical, but artifacts of self-reversed 
$I$ profiles.  The region is
entirely unipolar according to
the 854.2 nm data. 

The heated chromospheric and 30.4 nm
plasmas appear as rings  centered close to the coronal
loop footpoint emission. The lack of any correlation
between 30.4 and coronal emission indicates that 
this line of helium 
is more controlled by 
local plasma properties 
than  optically thin illumination
by radiation from the corona
\citep[a property noted by]{Judge+Pietarila2004}. 
Instead, given the 
ring of 30.4 emission appearing to surround the 17.1 emission, 
we can speculate that  the interface between the hot coronal loop's plasma 
and the surrounding 
plasma may play a role in 
generating extra heating
via collisional \textit{cross-field transport}
processes,  all within the same 
unipolar magnetic flux system. 
Cross-field processes have been previously studied
\citep[e.g.][]{Athay1990c,Ji+Song+Hu1996,
Judge+McIntosh2000, 
Ashbourn+Woods2001,Judge2008}.  
\citep{Ashbourn+Woods2001}
included a physical 
model for ion-acoustic turbulence to modify parallel and cross field transport.
Perhaps there is also a role for local but non-Maxwellian particle
distributions in the 
corona on scales of Mm, 
which are close to classical
mean free paths of ions and electrons. 
Interestingly,  the 30.4 nm
emission of Figure~\ref{fig:east} lies mostly \textit{in between} regions of
coronal emission, seen both as
loop-like emission and in individual patches
of 30.4 nm emission.  While differential projection effects cannot be ruled 
out to explain these offsets, the offsets are in random
directions which would be difficult to explain in
the relatively homogeneous chromospheric magnetic field beneath.

\section{Conclusions and further speculations}

The present analysis
suggests new avenues for making progress on a variety of puzzles involving 
magnetic plasma heating of the solar atmosphere. 
It may nevertheless appear 
superficially similar to earlier work,  recent examples being \citet{Chitta+others2017,Anan+others2021,EstabanPozuelo+others2023}.
We have taken special care to reduce and define uncertainties 
in co-alignment of diverse data across the entire atmosphere,  from
different instruments.  The whole
is greater than the sum of 
these components.

Our analysis also takes a  
broader view, deliberately stepping back to 
ask more elementary questions than is customary, based upon simple and obvious 
properties of the data, keeping 
firmly in mind the physical properties inferred from magnetic field
measurements and elementary magneto-hydrodynamics. 
We have striven to minimize confirmation bias \citep{BarkerBausell2021} by analyzing every pixel in all data the same way, using 
cross correlations and power spectra, without by-eye selection of 
identifiable 
phenomena.  
From
this more remote, and hopefully more objective    viewpoint, we have 
suggested that several 
assumptions concerning 
coronal heating should  be 
re-examined.

Uniquely, in stepping backwards and   seeking
to refute elementary predictions from  physical models,  we have shown that \textit{measurable} (i.e. large-scale) magnetic fields contain little information on 
plasma heating, by themselves. Hot coronal plasmas 
form as a result of both external
forcing \textit{and} internal dynamics \citep{Einaudi+others2021}. Thus, the bright coronal plasmas observed seem to remain so  as a result of ``hidden'', i.e.  unobserved processes, which are
internal to chromospheric and/or overlying plasmas themselves.   We argue that
the coronal plasma itself plays an active role in
the heating mechanisms,  chromospheric 
processes being of secondary importance.  
Theory would suggest that
such a non-linearity should be expected. Highlighted recently by \citet{Judge+Ionson2023}, this view is along
the same lines suggested independently by non-linear
MHD calculations
\citet{Einaudi+others2021}:
\begin{quote}
``Since the energy input is 
dependent on both the external 
forcing \textit{and the internal dynamics}, the 
corona is a
      self-regulating
forced system.''
(Our emphasis.)
\end{quote}
This idea is not new, it is found in early models of 1D flows 
along flux tubes, prompted by the seminal 
work of \citet{Rosner+Tucker+Vaiana1978}.  
\citet{Kuin+Martens1982,Martens+Kuin1983}
identified different attractor solutions, for the same physical ingredients,
in analytical/numerical dynamical models
with open connections 
to the chromosphere.   The present analysis
is perhaps \textit{the first to \textbf{require} multiple solutions
to explain the extreme variation of coronal heating, observed over 
a chromosphere whose measurable magnetic state is 
far more homogeneous}.    
This idea is also related to  
an earlier proposal for a ``critical level'' to be reached before significant 
coronal heating occurs  \citep{Litwin+Rosner1993}.  
A change in approaches to the coronal
heating problem to accommodate this result therefore appears 
necessary, if only to try to refute it. 

Our work begs further questions.  Where is the lower
temperature plasma emission arising from the conductive energy flux
down from the corona?  What is the physical nature of the plasmas
observed at transition region temperatures?  What is the role of
departures from Maxwellian distributions within and between ions and
electrons when mean free paths approach scales of macroscopic thermal
structures \citep{Jordan1975,Judge+Ionson2023}?  It appears that
whatever heats the 
bright EUV plasmas analyzed here must occur over locally unipolar
chromospheric fields, in which the measurable field plays only a minor
role in heating the plasma.  This conclusion is in line with a series
of arguments in the monograph by \citet{Judge+Ionson2023}. It
follows an earlier argument in a prescient article by
\citet{Sturrock1999}, that the magnetic fields \textit{on measurable
  scales} play a passive role in coronal heating, serving mostly to
set Alfv\'en speeds and topology more than determining plasma heating
rates.  If subsequently confirmed, this viewpoint
explains why 
earlier statistical
studies based upon observable 
magnetic fields ($\vec{B}_0$)
have been largely
inconclusive
\citep{Fisher+others1998,Mandrini+Demoulin+Klimchuk2000,Aschwanden+others2000}.

Lastly, we have inferred that bipolar fields measured in the photosphere have no role to play in heating the 
plasmas overlying the
particular observed regions. 
Based only on SDO AIA images and 
HMI longitudinal photospheric magnetic fields, 
\citet{Tiwari+others2021} recently also concluded that bipolar
fields are not \textit{necessary} for heating coronal loops. Our conclusions are stronger: such configurations are 
\textit{refuted} for the plasmas reported here. 

\appendix

\section{Cool loops: reductio ad absurdum}
\label{sec:noloops}

The cool loop proposal necessitates bipolar magnetic fields at scales
$\lesssim 10$ Mm
\citep{Dowdy+Rabin+Moore1986,Antiochos+Noci1986,Hansteen+others2014}.
Other models
\citep{Athay1990c,Ji+Song+Hu1996,Ashbourn+Woods2001,Ashbourn+Woods2006,Judge2008}
require no bipoles, but they invoked a variety of processes related
only to efficient ``classical'' rates of downward heat conduction
\citep{Spitzer1956,Braginskii1965}, and classical or turbulent
cross-field transport processes.  To survive scrutiny, the cool loop
model must satisfy a minimum of four requirements
\begin{enumerate}
    \item The loops must be observable at the UV and EUV  wavelengths at which
      transition region plasma is regularly observed. These include
      lines of the Li- and Na- isolelectronic sequences from
      \ion{C}{4} at 155 nm down to lines of Be-like \ion{Ne}{7} at
      46.5 nm, and \ion{Mg}{9} at 36.8 nm.
    \item Opposite polarity footpoints of bipoles must either be
      observed, or shown to be compatible with available Zeeman
      measurements.
    \item Loop-like structures between the footpoints must be
      responsible for the bulk of the transition region emission.
    \item The number and nature of such loops must be able to account
      for the brightness of lines such as H Ly-$\alpha$ and from ions
      typically two to four times ionized.
\end{enumerate}
The first requirement sets geometric constraints on the visibility of
cool loops.  Any proposed cool loops must extend about 0.8 Mm above
the quiet continuum photosphere in order to be visible at wavelengths
below 152 nm (\citealp{VALIII,Judge2015}, see also
figures 8--11 of \citealp{Skan+others2023}, showing calculations for
\ion{Si}{4} 139 and 140 nm lines).  Spectral lines below 110 nm can
only escape from regions 1.2 Mm above the photosphere. Smaller loops,
which certainly exist, are however irrelevant to the interpretation of
transition region emission.  If loops are close to a potential state
above the photosphere, they cannot extend much higher than the
separation between their footpoints.  MHD simulations of small loops
show that this argument is reasonable outside of highly dynamic
situations \citep{Skan+others2023}.  No matter the sensitivity or
angular resolution of any measurements, small bipoles with footpoints
separated by $\lesssim$ 0.8--1.2 Mm statistically cannot easily
contribute to observed cool loops.  So the only question remaining is
whether cool loops in the range of, say, 1--10 Mm can be consistent
with measurements of magnetic fields from ViSP.  The answer suggested
by our analysis is no.

The fourth requirement has been shown to be false by \citet{Judge2021}
in several quieter IRIS datasets obtained near disk center and at the
solar limb.  The third is evidently refuted by comparing the
top right images of photospheric $B_{LOS}$ measurements in
Figures~\ref{fig:east} and \ref{fig:west}, joining the opposite
polarity (white) patches to neighboring dominant polarities, and
looking to the bottom rows of both figures to see where emission
occurs.  There is no such correlation.
Our analysis reveals further problems with the notion of cool loops
for the bright network regions targeted by DKIST.  The domination of
one polarity over the $105\arcsec\times50\arcsec$ common field of
view, and its extension seen in HMI data, is extreme (compared with
lower resolution measurements discussed by \citealp{Giovanelli1980}):
97.5\%{} of the detectable magnetic flux is negative, 2.1\%{}
positive.  The upper right panels of Figures~\ref{fig:east} and
\ref{fig:west} show examples of small opposite-polarity flux
concentrations outside of the primary polarity.  The total flux of opposite polarity varies from the lower detection limit of $\approx 1.5\times10^{15}$
Mx in the 854.2 nm line (3 times the sensitivity level), up to the
largest observed fluxes of $10^{19}$ Mx.  \textit{If the ViSP has resolved all the
field present in the observations reported here} (i.e., there is no opposite polarity flux on scales below
$150$ km), then the measurements of $B_{LOS}$ in the 630.2 and 854.2
nm lines in these figures reveal the sheer impossibility that loops
returning within the vast, fairly uniform unipolar regions seen in the
854.2 magnetograms can exist.  However, is it possible to
hide opposite polarity flux in a physically meaningful manner?

It seems only one option is left to ``save the phenomenon'' of the
cool loop picture. To account for the brightness of plasmas above the
dominant unipolar network concentrations, we must hide tubes of
magnetic flux beneath detection levels of $\approx 1.5\times10^{15}$ Mx with
footpoints that must be separated by at least 1 Mm ($1\farcs4$). This means
we must seek in the 854.2 magnetograms, formed 1400 km above the
photosphere, signals of such concentrations which are associated with
one end of a cool loop (the other being the dominant majority
polarity). Close inspection of the figures show no significant
correlation of this kind.

We can also bring a physical argument that the field
\textit{strengths} above a certain height in the chromosphere must be
about the same magnitude owing to
horizontal force balance. The measured unipolar flux densities in the 854.2 nm
line generally exceed $B > 100$ Mx~cm$^{-2}$ over the network. Using the 
pressure at 1400 km from
model C of \cite{VALIII},
the
plasma $\beta=8\pi p/B^2$ must satisfy $ \beta < 0.005 $. The associated Alfv\'en speed
is $\gtrapprox 500$ km~s$^{-1}$.  Thus  any magnetic
field in the chromosphere no matter its sign, will quickly reach a field strength of at least $100$ G, because 
no force can stop the magnetic field 
from filling space.  With $B > 100$ G, flux $< 2\times10^{15}$
Mx, the tube area would not exceed $2\times10^{13}$ cm$^2$. A cylindrical tube    with this area would have a radius 25 km. If we add in 
one more constraint from
observations, we will
find bizarre unphysical 
properties of such cool loops.  The emission measure in the lower transition region, averaged
over much larger areas, is roughly $10^{27}$ cm$^{-5}$
\citep{Judge+others1995}. That is, when averaged over areas $S$ of
several Mm$^2$, from which emission measures are derived from earlier
observations, the average of $\int n_e^2 dz$ is $\approx 10^{27}$
cm$^{-5}$.  Assuming that this average applies roughly to the
$105\arcsec\times50\arcsec$ area observed by DKIST (an area of
$3\times10^{19}$ cm$^2$, see Figure~\ref{fig:fov}), we would require an
average volumetric emission measure of $S \int n_e^2 z \approx
3\times10^{46}$ cm$^{-3}$.  With $n_e \approx 10^{10}$ cm$^{-3}$ a typical
estimate of electron density, the emitting plasmas must fill a volume
$V \approx 3\times10^{26}$ cm$^3$.  If this is to be provided by cool loops
of radius $< 2.5\times10^6$ cm, the total length of cool loops would need
to exceed $1.5\times10^{13}$ cm, over 210 solar radii or about 1
astronomical unit.  Another way of saying this is that for each loop
limited to being less than 10 Mm in order for them to be confined near
the network boundaries, we would need $>1.5\times10^7$ loops scattered
across the DKIST common field of view of $3\times10^{19}$ cm$^2$.  If we
further confine these to live within the network boundaries this would
imply approximately 1 cool loop, of radius 25 km, every 10 meters
across the length of network boundaries!

\section*{Acknowledgments}
This material is based upon work supported by the National Center for
Atmospheric Research, which is a major facility sponsored by the National Science Foundation under 
Cooperative Agreement No. 1852977.  PGJ is grateful to HAO 
Director Holly Gilbert for support of a research visit in 
Switzerland in 2023, to Daniela Lacatus for a thorough 
review of the article, and to 
Christian Beck and Gianna Cauzzi
for their work acquiring the
DKIST data and useful comments.
Both the University of Bern and the 
International Space Science Institute  supported this work, 
and grant No. 216870 from the Swiss Science Foundation to LK made this collaboration 
possible. The research reported herein is based in part on data 
collected with the Daniel K. Inouye Solar Telescope (DKIST), a facility of the National Solar Observatory (NSO). NSO is managed by the Association of Universities for Research in Astronomy, Inc., and is funded by the National Science Foundation. Any opinions, findings and conclusions or recommendations expressed in this publication are those of the authors and do not necessarily reflect the views of the National Science Foundation or the Association of Universities for Research in Astronomy, Inc. DKIST is located on land of spiritual and cultural significance to Native Hawaiian people. The use of this important site to further scientific knowledge is done so with appreciation and respect.

\bibliographystyle{aasjournal} \bibliography{best}
\end{document}